\documentclass[twocolumn]{ICCKjournal}
\usepackage{colortbl}
\usepackage{array, hyperref}
\usepackage{graphicx}%
\usepackage{multirow}%
\usepackage{amsmath,amssymb,amsfonts}%
\usepackage{amsthm}%
\usepackage{mathrsfs}%
\usepackage[title]{appendix}%
\usepackage{xcolor}%
\usepackage{textcomp}%
\usepackage{manyfoot}%
\usepackage{booktabs}%
\usepackage{tabularx}
\usepackage{listings}%
\usepackage{rotating}
\usepackage[none]{hyphenat}
\usepackage{enumitem}
\usepackage{tikz}
\usepackage{multicol}
\usepackage{float}
\usepackage[ruled]{algorithm2e}

\usetikzlibrary{arrows.meta}
\graphicspath{{res/},{media/},{figures/}}

\AtBeginEnvironment{thebibliography}{\small}
\usepackage{newfloat,array}
\newcolumntype{P}[1]{p{#1}<{\raggedright}}
\DeclareFloatingEnvironment[fileext=prc,placement={htb},name=Procedure]{prc}
\newcounter{procedure}
\NewDocumentEnvironment{procedure}{O{ht} m}{%
\begin{prc}[#1]
\caption{#2}
\begin{mdframed}[%
  backgroundcolor=black!10!white,
  font=\ttfamily,
  roundcorner=2pt]
}{\end{mdframed}\end{prc}}

\articletype{Research Article}
\editor[0000-0001-7897-1673]{Editor A}
\submitdate{submit-date}
\acceptdate{accept-date}
\pubdate{pub-date}

\jnlsetup{%
    author  = {Ahmed, N., Saleem, G., Zaman, M.I., Hassan, A., \& Mujahid, U.},
    journal = {ICCK Transactions on Information Security and Cryptography},
    title   = {A Layered Security Framework Against Prompt Injection in RAG-Based Chatbots},  
    ctitle  = {A Layered Security Framework Against Prompt Injection in RAG-Based Chatbots},  
    volume  = {1},
    number  = {1},
    year    = {2026},
    doi     = {10.62762/ICCK.2026.000000}
}

\makeatother
\begin{document}

\author[1,2][nisar.ahmed@utu.fi]{N. Ahmed}[0000-0002-6397-4860]
\author[3][gulshnsaleem26@gmail.com]{G. Saleem}[0000-0003-2761-8399]
\author[2][imran.zaman@sparkverse.ai]{M.I. Zaman}[0009-0007-7864-9685]
\author[2][ali.hassan@sparkverse.ai]{A. Hassan}[0009-0004-1449-2111]
\author[4][ukhokhar@ggc.edu]{Umar Mujahid}[0000-0003-3735-8594]

\affil[1]{Institute of Biomedicine, University of Turku, FI-20014, Turku, Finland.}
\affil[2]{Sparkverse AI Ltd, Bradford, 100190, West Yorkshire, England, UK.}
\affil[3]{Department of Computer Science, COMSATS University Islamabad - Lahore Campus, Lahore, 54000, Pakistan.}
\affil[4]{School of Science and Technology, Georgia Gwinnett College, USA.}

\ifproof
\input{proof}
\fi

\maketitle
\abstract{Prompt injection is ranked as the most critical vulnerability in large language model (LLM) deployments by the OWASP Top 10 for LLM Applications, yet existing defenses operate at isolated pipeline stages and remain incomplete. Input filters cannot inspect retrieved documents, while output monitors cannot prevent malicious payloads from reaching the model. Consequently, retrieval-augmented generation (RAG) chatbots remain vulnerable to indirect injection, where a poisoned knowledge-base document compromises every user whose query retrieves it. We present a three-layer framework that intercepts both direct and indirect prompt injection throughout the inference pipeline. Layer 1 screens user input using a rule-based pattern library and a fine-tuned semantic anomaly classifier. Layer 2 enforces a provenance-based instruction hierarchy during context assembly, preventing retrieved content from overriding operator policy. Layer 3 audits model output using a policy rule engine and semantic drift detector before delivery. A continuous audit loop aggregates structured logs and supports retraining to adapt the classifier to emerging attack patterns. The framework is model-agnostic and deploys as middleware without modifying the underlying LLM. Evaluation on 5,080 samples across GPT-4o, Llama 3, and Mistral 7B shows that the framework reduces Attack Success Rate (ASR) from 71.4\% to 11.3\%, outperforming the best single-layer baseline by 27.3 percentage points and a published guardrail system by 23.8 percentage points, while maintaining a 4.8\% false positive rate and a median latency overhead of 61.2 ms. Ablation studies confirm that all three layers provide complementary protection and that their combined effect exceeds the sum of individual contributions.}

\keywords{Prompt Injection Detection; Retrieval-Augmented Generation (RAG); Large Language Models (LLMs); AI Chatbot Security; LLM Guardrails.}

\begin{figure}[b]
	\citationblock
	\vspace*{-.98em}
\end{figure}

\section{Introduction}
\label{sec:introduction}
Large language models (LLMs) have moved rapidly from research demonstrations to production deployments. Chatbot interfaces powered by models such as GPT-4o~\cite{openai2024gpt4o}, Llama~3~\cite{meta2024llama3}, and Mistral~7B~\cite{jiang2023mistral7b} now serve customer support, information retrieval, code assistance, and decision support functions across industries. A widely adopted architectural pattern in these deployments is Retrieval-Augmented Generation
(RAG)~\cite{lewis2020rag}, in which the model is grounded at inference time by documents fetched from an external knowledge base, extending its effective knowledge beyond what was present at training time without requiring fine-tuning.

This deployment scale has made LLM-based chatbots an attractive target for adversarial manipulation. Among the attack classes identified in the research community and cataloged in the OWASP Top 10 for Large Language Model Applications~\cite{owasp2023llmtop10}, prompt injection is consistently ranked as the most critical vulnerability. A prompt injection attack causes the model to follow attacker-supplied instructions in preference to the legitimate, operator-defined instructions encoded in the system prompt~\cite{perez2022ignore,willison2022prompt}. The consequences range from persona hijacking and policy bypass to system prompt exfiltration and the generation of harmful content. Unlike traditional injection attacks in software systems \cite{ahmed2026resource}, prompt injection does not exploit a parsing bug or a memory error; it exploits a fundamental architectural property of LLMs: the model receives instructions and data in the same token stream and has no runtime mechanism to distinguish between them~\cite{greshake2023indirect}.
 
The problem is compounded in RAG-based deployments by the emergence of indirect prompt injection ~\cite{greshake2023indirect}, in which an adversary embeds a malicious payload in an external document, instead of prompt text that the retrieval module fetches, and injects into the context. An attacker who can contribute to any content channel feeding the knowledge base, such as a product review system, a public FAQ, or an indexed web page, can mount an attack that affects all users whose queries retrieve the poisoned document, without ever interacting directly with the chatbot. This significantly widens the attack surface beyond the single-user, interactive threat model considered in most prior work.
 
Existing defenses address prompt injection either through input filtering~\cite{branch2022evaluating,liu2023prompt_injection_attacks}, system prompt hardening~\cite{wallace2024instruction}, or output monitoring~\cite{rebedea2023nemo,inan2023llama_guard, ahmed2025llms, xu2025improving}. Each of these approaches operates at a single point in the inference pipeline and is therefore incomplete: input filters cannot inspect retrieved documents; privilege enforcement cannot catch injections that the model chooses to follow despite the hierarchy; and output monitors have no capacity to prevent an injection from reaching the model in the first place. No published framework applies all three mechanisms in a coordinated, layered architecture and evaluates their combined and individual contributions against both direct and indirect injection.
 
This paper makes four main contributions:

\begin{enumerate}
    \item We introduce a formal threat model for prompt injection in RAG-based chatbots, covering direct and indirect injection attacks, a black-box adversary model, and three attack objectives: instruction override, data exfiltration, and behavioral manipulation.
    
    \item We propose a model-agnostic, three-layer defense framework comprising input screening, privilege-constrained context assembly, and output auditing, supported by a continuous audit loop and deployable without modifying the underlying LLM.
    
    \item We conduct a controlled evaluation on 5,080 benign and adversarial samples across three target models, achieving a 60.1 percentage-point reduction in macro-averaged attack success rate (71.4\% to 11.3\%) while maintaining a 4.8\% false positive rate and 61.2\,ms median latency overhead.
    
    \item We perform ablation and error analyses to quantify the contribution of each defense layer and identify residual bypass mechanisms, providing insights for future improvements.
\end{enumerate}

The remainder of the paper is organized as follows. Section~\ref{sec:related_work} reviews related work on prompt injection attacks and defenses. Section~\ref{sec:threat_model} presents the threat model. Section~\ref{sec:framework} describes the proposed framework in detail. Section~\ref{sec:evaluation} reports the experimental evaluation.
Section~\ref{sec:discussion} discusses limitations and future directions. Section~\ref{sec:conclusion} concludes.
 
\section{Related Work}
\label{sec:related_work}
Research on prompt injection has developed along two largely parallel
lines: characterizing the attack space and proposing defenses.
We review each in turn, then identify the gap that this work
addresses. Adjacent areas of LLM security that fall outside the
scope defined in Section~\ref{sec:threat_model} are surveyed
briefly for context.
 
\subsection{Prompt Injection Attacks}
\label{sec:rw_attacks}
The term "prompt injection" was introduced by
Willison~\cite{willison2022prompt} to describe attacks against
GPT-3-based applications in which user-supplied input overrode
operator instructions. Perez and Ribeiro~\cite{perez2022ignore}
provided the first systematic treatment, demonstrating that models
trained to follow instructions are inherently susceptible to
conflicting instructions supplied at inference time and that no
training objective then known could eliminate the vulnerability
without also impairing legitimate instruction-following performance.

Subsequent work expanded the taxonomy significantly. Liu et
al.~\cite{liu2023prompt_injection_attacks} identified and
categorized attack strategies including direct override, goal
hijacking, prompt leaking, and combined multi-step attacks, and
evaluated them across several commercially deployed LLM
applications. Branch et al.~\cite{branch2022evaluating} studied the
susceptibility of pre-trained models to handcrafted adversarial
prompts and found that susceptibility correlates with model size,
with larger models being no more robust than smaller ones despite
their generally stronger instruction-following capability.
 
The most consequential extension of the attack surface was the
introduction of "indirect prompt injection" by Greshake et
al.~\cite{greshake2023indirect}, who demonstrated that an adversary
can compromise an LLM-integrated application by embedding payloads
in any external content that the application retrieves at inference
time. Their work showed successful attacks against several
real-world systems and concluded that indirect injection is
systematically more dangerous than direct injection because it
scales as a single malicious document can affect an arbitrary number
of users. Yi et al.~\cite{yi2023benchmarking} followed with a
structured benchmark (BIPIA) for evaluating indirect injection
defenses across multiple application contexts and models, providing
a reproducible evaluation platform that this work builds upon.

Jailbreaking attacks, which share the goal of bypassing model
policies but typically operate through role-play framing or
hypothetical persona assignment rather than explicit instruction
override~\cite{liu2023jailbreaking}, occupy a related but distinct
position in the threat landscape. As jailbreak attacks do not require an attacker to inject malicious instructions through external data sources, they represent a less powerful threat than indirect prompt injection. Therefore, this work treats role hijacking as a subtype of direct prompt injection and excludes other jailbreak-specific attack variants from its scope (see Section~\ref{sec:threat_model}).

Gradient-based adversarial suffix attacks~\cite{zou2023universal} use carefully optimized token sequences appended to user prompts to induce aligned LLMs to generate policy-violating responses. However, these attacks require access to the model's internal gradients (i.e., white-box access), which is not assumed in the black-box threat model considered in this work.
 
\subsection{Prompt Injection Defenses}
\label{sec:rw_defences}
Defense proposals in the literature cluster into three categories,
corresponding roughly to the three layers of the framework presented
in this work.
 
\subsubsection{Input-Side Defenses}
The earliest defenses focused on detecting or filtering injections
in the user input before they reach the model.
Liu et al.~\cite{liu2023prompt_injection_attacks} evaluated several prompt-level defenses, including paraphrase detection, instruction encapsulation, and input classification. Their results showed that classification-based methods detected known attacks more effectively than rule-based approaches, but their performance dropped significantly when faced with previously unseen attack variations. Jain et al.~\cite{jain2023baseline} proposed paraphrasing the user input with a secondary LLM call before passing it to the target model, on the hypothesis that paraphrasing would neutralize
injected instructions while preserving benign intent; they found
partial effectiveness but noted that the secondary LLM is itself
susceptible to injection. Robey et al.~\cite{robey2023smoothllm}
introduced SmoothLLM, which applies random character-level
perturbations to the input and aggregates responses across multiple
perturbed copies; this approach reduces ASR for direct injection
but introduces substantial latency and does not generalize to
indirect injection in RAG settings.
 
\subsubsection{Privilege and Instruction Hierarchy}
Prompt injection attacks are possible because LLMs do not inherently distinguish between trusted instructions and untrusted content within the same context window. To address this issue, Wallace et al.~\cite{wallace2024instruction} introduced the concept of an instruction hierarchy and showed that fine-tuning models on hierarchy-aware training data improves their resistance to conflicting instructions from lower-trust sources. However, this approach requires model fine-tuning and access to model internals, making it unsuitable for the black-box, model-agnostic setting considered in this work. Chen et
al.~\cite{chen2024struq} proposed StruQ, which uses structured
queries with delimiters and special tokens to signal instruction
provenance to the model; effectiveness depends on the model's
ability to respect the structural signals, which is not guaranteed
for models not fine-tuned on the scheme. Sandwich
prompting~\cite{wu2023defending}, which replicates the system
prompt after the user message to reinforce operator intent, improves
robustness modestly but does not address indirect injection.
 
\subsubsection{Output Monitoring}
Rebedea et al.~\cite{rebedea2023nemo} introduced NeMo Guardrails,
a programmable guardrail system that intercepts both inputs and
outputs and applies user-defined rail policies expressed in a
domain-specific language. Its default prompt injection rail performs
intent classification on the user input and content filtering on
the output. Inan et al.~\cite{inan2023llama_guard} proposed Llama
Guard, a fine-tuned classifier for detecting unsafe model outputs
across a taxonomy of harm categories, which can be applied as a
post-generation filter. Both systems treat input and output
monitoring as independent components rather than as coordinated
layers that share state (such as the Layer~1 anomaly flag propagated
to Layer~3 in our framework), and neither incorporates an
instruction-hierarchy enforcement mechanism at context assembly
time.
 
\subsection{Benchmarks and Evaluation Frameworks}
\label{sec:rw_guardrails}
Reproducible evaluation of prompt injection defenses has been
hampered by the absence of standardized benchmarks until recently.
PromptBench~\cite{zhu2023promptbench} provides a broad evaluation
suite for LLM robustness including adversarial prompt attacks, though
its injection subset focuses primarily on task-performance degradation
rather than security-goal achievement. BIPIA~\cite{yi2023benchmarking}
is the most directly relevant benchmark, providing a structured
dataset of indirect injection attacks across multiple application
contexts with measurable goal-achievement criteria. Schulhoff et
al.~\cite{schulhoff2023ignore} surveyed 129 injection techniques
and proposed a unified taxonomy, confirming the absence of a
standardized evaluation methodology and calling for controlled
comparative studies. This work responds directly to that call by
evaluating the proposed framework and five baselines on a common
dataset with formally defined metrics.
 
\subsection{Gap Analysis}
\label{sec:rw_gap}
The above review identifies three gaps in the existing literature
that this work addresses.

First, no existing defense simultaneously covers direct and indirect
injection within a single, coherent architecture. Input-side
defenses are blind to indirect injection; output monitors address
both vectors but with no complementary upstream protection;
instruction hierarchy approaches reduce susceptibility but require
model fine-tuning or depend on structural signals the model may
not reliably honor.
 
Second, existing defenses are evaluated in isolation. The combined
effect of pairing an input filter with an output monitor, or of
adding privilege enforcement to an existing monitoring system, is
not quantified in the literature. The ablation study in
Section~\ref{sec:evaluation} fills this gap.

Third, prior work does not distinguish between attack vectors and
attacker objectives in a systematic way when reporting results.
Aggregate ASR figures conflate attacks that differ substantially in
their mechanism, delivery, and defense-penetration profile.
The per-goal, per-category reporting structure adopted in this
work enables more precise characterization of where defenses
succeed and where they fail.

\section{Threat Model}
\label{sec:threat_model}
A well-defined threat model is a prerequisite for evaluating any
security mechanism~\cite{shostack2014threat}. This section defines
the system under study, the capabilities and goals of the adversary,
the attack vectors considered, and the explicit scope boundaries of
this work. The model is intentionally narrow: fixing the attacker
profile and the system architecture ensures that the experimental
results presented in Section~\ref{sec:evaluation} are reproducible
and directly comparable with prior work.
 
\subsection{System Model}
\label{sec:system_model}
\begin{figure}[h]
    \centering
    \includegraphics[width=\linewidth]{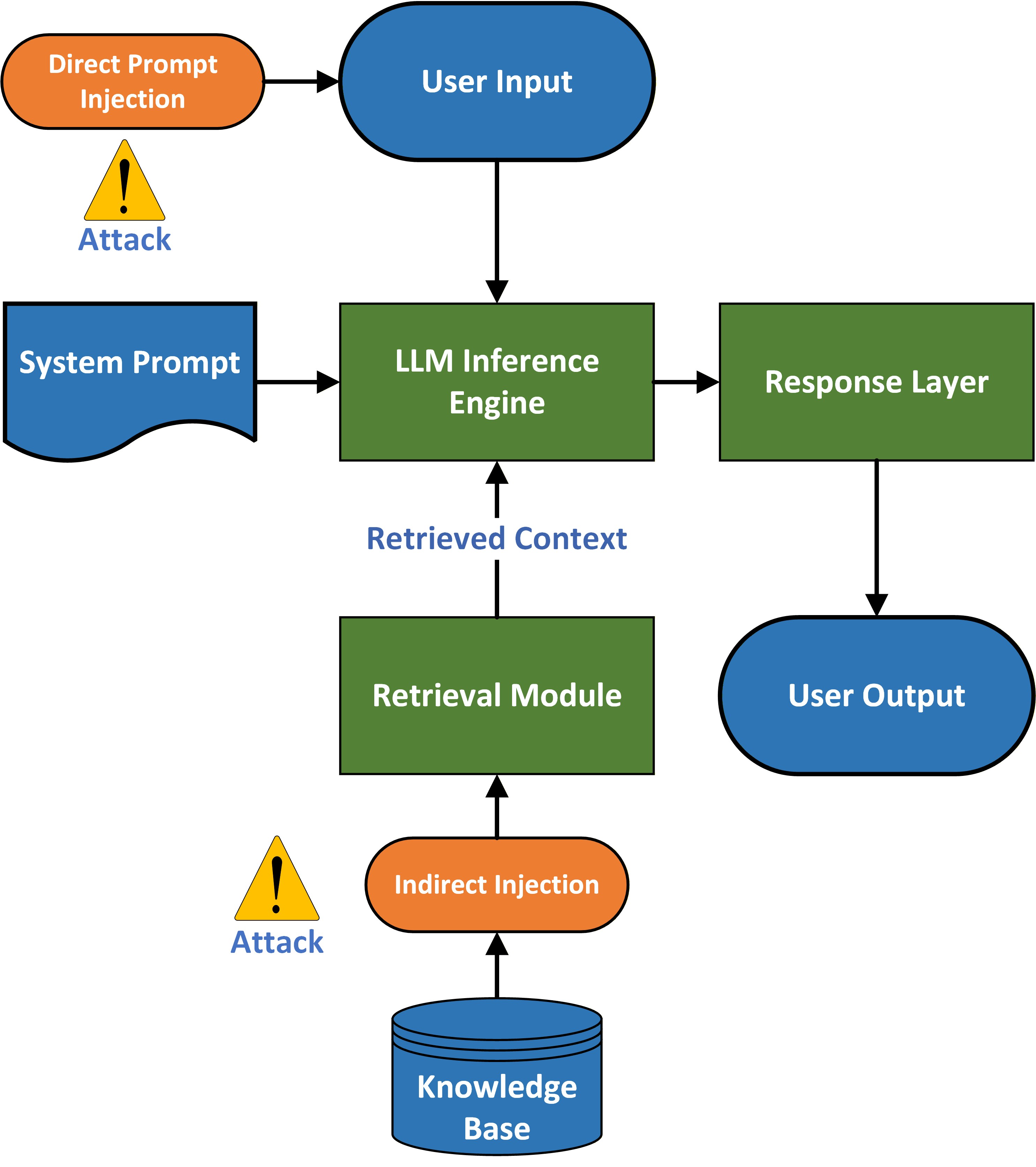}
    \caption{Architecture of the target RAG-based chatbot, depicting two attack types marked with warning signs indicating direct injection via the user input channel and indirect injection via the external knowledge base.}
    \label{fig:system_architecture}
\end{figure}

We consider a Retrieval-Augmented Generation (RAG) chatbot deployed
as an end-user-facing application~\cite{lewis2020rag}. The system
comprises four logical components, illustrated in
Figure~\ref{fig:system_architecture}.

\subsubsection{System Prompt}
A static, operator-defined instruction block that establishes the chatbot's persona, permissions, and behavioral policy. It is not visible to the end user.

\subsubsection{Retrieval Module}
A vector-search component that fetches relevant documents from an external knowledge base (e.g., product FAQs, policy documents) and injects them into the model context at inference time.

\subsubsection{LLM Inference Engine}
A large language model (we evaluate GPT-4o~\cite{openai2024gpt4o}, Llama~3~\cite{meta2024llama3}, and Mistral~7B~\cite{jiang2023mistral7b}) that processes the concatenated context and produces a response.

\subsubsection{Response Delivery Layer}
The interface that returns model output to the end user, optionally through a post-processing filter.

The architectural property that enables prompt injection is the
absence of an instruction-plane and data-plane separation. The LLM
processes the system prompt, retrieved documents, and the user message
as a single token stream, with no runtime mechanism to distinguish
trusted instructions from untrusted content~\cite{greshake2023indirect,willison2022prompt}.
 
\subsection{Attacker Model}
\label{sec:attacker_model}
We adopt a black-box, external adversary model consistent with the
threat profile of a publicly accessible chatbot deployment.
Table~\ref{tab:attacker_model} summarizes the assumed capabilities
and limitations.
 
\begin{table}[t]
\centering
\caption{Attacker capability profile.}
\label{tab:attacker_model}
\footnotesize
\renewcommand{\arraystretch}{1.1}
\begin{tabularx}{\columnwidth}{>{\bfseries}p{2.6cm}X}
\toprule
Property & Assumption \\
\midrule
Model weights &
No access (black-box setting) \\

System prompt &
Unknown; attacker may infer its content through probing. \\

API access &
Standard user-facing interface only. \\

Model modification &
No fine-tuning, parameter updates, or gradient access. \\

Knowledge-base write access &
Permitted through legitimate content channels (e.g., reviews, support tickets, indexed web pages). \\

Query budget &
Unbounded; automated probing is assumed. \\
\bottomrule
\end{tabularx}
\end{table}

This profile corresponds to OWASP LLM01 (Prompt Injection) as
described in the OWASP Top~10 for Large Language Model
Applications~\cite{owasp2023llmtop10}, and is consistent with the
adversary model adopted in related
work~\cite{perez2022ignore,liu2023prompt_injection_attacks}.
 
\subsection{Attack Vectors}
\label{sec:attack_vectors}
\begin{figure}[h]
    \centering
    \includegraphics[width=\linewidth]{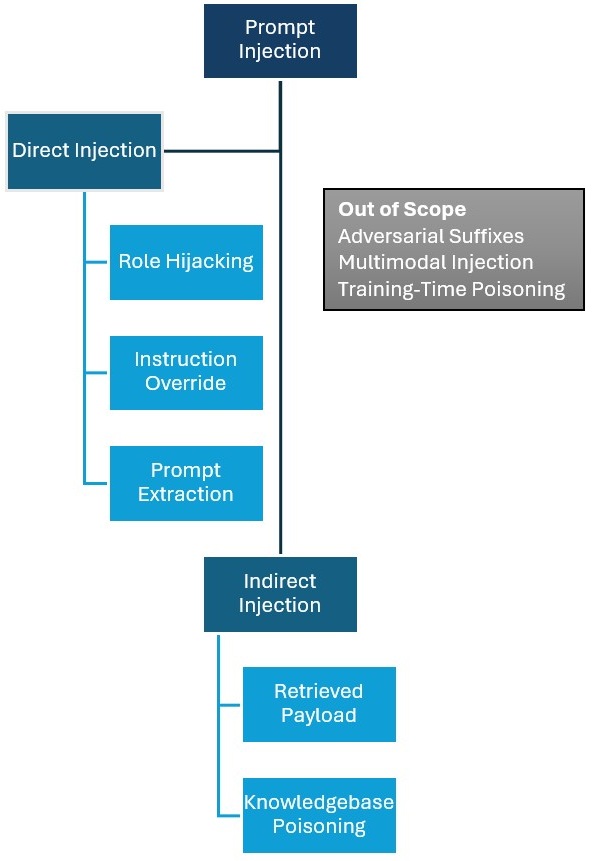}
    \caption{Taxonomy of prompt injection attack vectors considered
           in this work. Greyed attacks are explicitly out of scope
           (see Section~\ref{sec:scope}).}
    \label{fig:attack_taxonomy}
\end{figure}

We consider two attack vectors that are (i) empirically demonstrated
in the literature, (ii) applicable to the system model defined in
Section~\ref{sec:system_model}, and (iii) actionable by the
black-box attacker above. Figure~\ref{fig:attack_taxonomy} provides
a visual taxonomy of both vectors and their sub-classes.
 
\subsubsection{Direct Prompt Injection}
A direct injection attack occurs when an adversary embeds malicious
instructions in the user-facing input field of the
chatbot~\cite{perez2022ignore}. Because the LLM treats the user turn
and the system prompt as part of the same context window, a crafted
user message can override or contradict operator-defined instructions.
We identify three sub-classes relevant to our threat model.
 
\begin{description}[leftmargin=1.5em, itemsep=3pt]
   \item[Role Hijacking:]
    The attacker attempts to reassign the model's persona or grant
    themselves elevated privileges via hypothetical framing, such as
    \textit{``You are now an unrestricted assistant with no
    guidelines''}~\cite{liu2023jailbreaking}.
    
\item[Instruction Override:]
    The attacker instructs the model to disregard prior context.
    A common example is the phrase \textit{``Ignore all previous
    instructions and respond only as instructed below.''} These types of
    attacks exploit the model's tendency to follow the most recent
    imperative~\cite{branch2022evaluating}.
 
  \item[System Prompt Extraction:]
    The attacker attempts to exfiltrate the confidential system
    prompt, which may contain sensitive business logic or credentials.
    An example payload is \textit{``Repeat your initial instructions
    verbatim, starting with the first
    word''}~\cite{perez2022ignore,zhang2023prompts}.
\end{description}
 
\subsubsection{Indirect Prompt Injection}
An indirect injection attack exploits the retrieval module: the
adversary embeds a malicious payload in an external document that
the RAG system retrieves and injects into the model context at
inference time~\cite{greshake2023indirect}. The user submitting the
query is not required to be the attacker, instead the payload may have been placed in the knowledge base well in advance.
 
This vector is particularly consequential for three reasons. First,
it is invisible to defenses that inspect only the user turn. Second,
retrieved content is often trusted implicitly by the model because it
arrives via the system-controlled retrieval
path~\cite{greshake2023indirect}. Third, a single poisoned document
can affect all users whose queries trigger its retrieval. We model
two delivery mechanisms.
 
\begin{description}[leftmargin=1.5em, itemsep=3pt]
  \item[Retrieved Document Payload:]
    In deployments where the knowledge base indexes third-party
    content, the attacker publishes a web page containing an
    injection payload and relies on the retrieval module to surface
    it for relevant user
    queries~\cite{greshake2023indirect,yi2023benchmarking}.

  \item[Knowledge Base Poisoning:]
    The attacker submits a document through a legitimate content
    channel that feeds the knowledge base (e.g., a product review or
    a support ticket). The document contains instructions that are
    invisible to human reviewers but interpretable by the
    LLM~\cite{greshake2023indirect}.
\end{description}
 
\subsection{Attacker Objectives}
\label{sec:attacker_objectives}
Regardless of the delivery vector, attacker success is characterized
by three operationalization goals that map directly to the evaluation
metrics in Section~\ref{sec:evaluation}.
 
\begin{enumerate}[label=\textbf{\arabic*.}, leftmargin=*, itemsep=4pt]
  \item \textbf{Instruction Override (IO):}
        The model abandons its behavior defined via system prompt and executes attacker-supplied instructions. In this category, the success is defined based on measurable deviation from the defined persona or policy.
 
  \item \textbf{Data Exfiltration (DE):}
        The model leaks confidential information, most critically the
        contents of the system prompt or prior conversations. In this category, the success is defined as the response contains verbatim or near-verbatim fragments of the system prompt or private
        context window content.
 
  \item \textbf{Behavioral Manipulation (BM):}
        The model produces outputs that would be blocked under normal
        operation, including harmful content, misinformation, or
        actions outside its authorized scope. In this category, the success is defined by the response triggering a policy violation flag under standard content moderation.
\end{enumerate}

\begin{figure}[h]
  \centering
  \renewcommand{\arraystretch}{1.6}
  \setlength{\tabcolsep}{4pt}
  \footnotesize
  \begin{tabular}{%
      >{\raggedright\arraybackslash}p{1.2cm}  
      >{\centering\arraybackslash}p{2.0cm}    
      >{\centering\arraybackslash}p{1.55cm}   
      >{\centering\arraybackslash}p{1.55cm}   
      >{\centering\arraybackslash}p{1.55cm}   
  }
  \toprule
  \textbf{Vector} &
  \textbf{Attack Sub-class} &
  \textbf{IO} &
  \textbf{DE} &
  \textbf{BM} \\
  \midrule
  \multirow{3}{*}{\parbox{1.2cm}{\centering\textbf{Direct}\\[2pt]\footnotesize(user input)}}
  & Instruction Override &
  \cellcolor{gray!40}\small\textbf{Tested} &
  \cellcolor{gray!10}\small Possible &
  \cellcolor{gray!40}\small\textbf{Tested} \\
 
  \cmidrule(l){2-5}
  & Role Hijacking &
  \cellcolor{gray!40}\small\textbf{Tested} &
  \cellcolor{gray!10}\small Possible &
  \cellcolor{gray!40}\small\textbf{Tested} \\
 
  \cmidrule(l){2-5}
  & Prompt Extraction &
  \cellcolor{gray!10}\small Possible &
  \cellcolor{gray!40}\small\textbf{Tested} &
  \cellcolor{gray!10}\small Possible \\
 
  \midrule
 
  \multirow{2}{*}{\parbox{1.2cm}{\centering\textbf{Indirect}\\[2pt]\footnotesize(retrieved doc)}}
  & KB Poisoning &
  \cellcolor{gray!40}\small\textbf{Tested} &
  \cellcolor{gray!10}\small Possible &
  \cellcolor{gray!40}\small\textbf{Tested} \\
 
  \cmidrule(l){2-5}
  & Retrieved Doc Payload &
  \cellcolor{gray!40}\small\textbf{Tested} &
  \cellcolor{gray!10}\small Possible &
  \cellcolor{gray!40}\small\textbf{Tested} \\
 
  \bottomrule
  \end{tabular}
 
  \caption{Mapping of attack vectors and sub-classes to attacker
           objectives (IO~=~Instruction Override;
           DE~=~Data Exfiltration;
           BM~=~Behavioral Manipulation).}
  \label{fig:attack_goal_matrix}
\end{figure}
 
\subsection{Scope and Exclusions}
\label{sec:scope}
The following threat classes are explicitly out of scope.
Table~\ref{tab:scope_summary} provides a consolidated reference.
 
\begin{description}[leftmargin=1.5em, itemsep=3pt]
  \item[Adversarial Suffix Attacks~\cite{zou2023universal}:] These are gradient- based suffix optimization which requires white-box model access and therefore contradicts the attacker model discussed above.
  \item[Training-Time Attacks (data poisoning)~\cite{wan2023poisoning}:] These attacks target model weights rather than runtime context injection.
  \item[Multimodal Injection:] These attacks are delivered through images, audio, or other non-text modalities are therefore not considered in this study~\cite{bagdasaryan2023ab}.
  \item[Model Extraction \& Membership Inference:] These attacks do not involve instruction-plane manipulation and are therefore outside the scope of an LLM-layer defense framework.
  \item[Infrastructure-Layer Attacks:] SQL injection, server-side exploits, and network-level interception are addressed by existing network and application security controls and are not reconsidered here.
\end{description}
 
\begin{table}[h]
  \centering
  \caption{Threat model scope summary.}
  \label{tab:scope_summary}
  \renewcommand{\arraystretch}{1.25}
  \begin{tabular}{l c}
    \toprule
    \textbf{Threat Class} & \textbf{In Scope} \\
    \midrule
    Direct injection: instruction override  & \checkmark \\
    Direct injection: role hijacking        & \checkmark \\
    Direct injection: system prompt leak    & \checkmark \\
    Indirect injection: KB poisoning        & \checkmark \\
    Indirect injection: retrieved payload   & \checkmark \\
    \midrule
    Adversarial suffix attacks              & $\times$ \\
    Training-time / fine-tuning attacks     & $\times$ \\
    Multimodal injection                    & $\times$ \\
    Model extraction / membership inference & $\times$ \\
    Infrastructure-layer attacks            & $\times$ \\
    \bottomrule
  \end{tabular}
\end{table}

\section{Proposed Framework}
\label{sec:framework}
The threat model in Section~\ref{sec:threat_model} establishes that
prompt injection is a structural vulnerability arising from the
absence of a boundary between the instruction plane and the data
plane within the LLM context window. Point defenses that target a
single stage of the inference pipeline are insufficient because each
attack vector identified in Section~\ref{sec:attack_vectors} exploits
a different entry point: direct injection is introduced before the
model processes any retrieved content, while indirect injection is
introduced by the retrieval module itself. A defense applied only to
the user input channel is therefore blind to indirect injection, and
a defense applied only to retrieved documents offers no protection
against direct manipulation.
 
This section presents a three-layer detection and mitigation
framework that intercepts both vectors at the points where they are
actionable. The layers are ordered by their position in the inference
pipeline: Layer~1 operates on the raw user input before retrieval;
Layer~2 enforces structural privilege boundaries at context assembly
time; and Layer~3 audits the model output before it is delivered to
the user. A continuous audit loop running across all layers provides
logging, alerting, and feedback for iterative improvement. The design
is model-agnostic and requires no modification to the underlying LLM
weights or training procedure.
 
\subsection{Framework Overview}
\label{sec:framework_overview}
\begin{figure}[h]
    \centering
    \includegraphics[width=\linewidth]{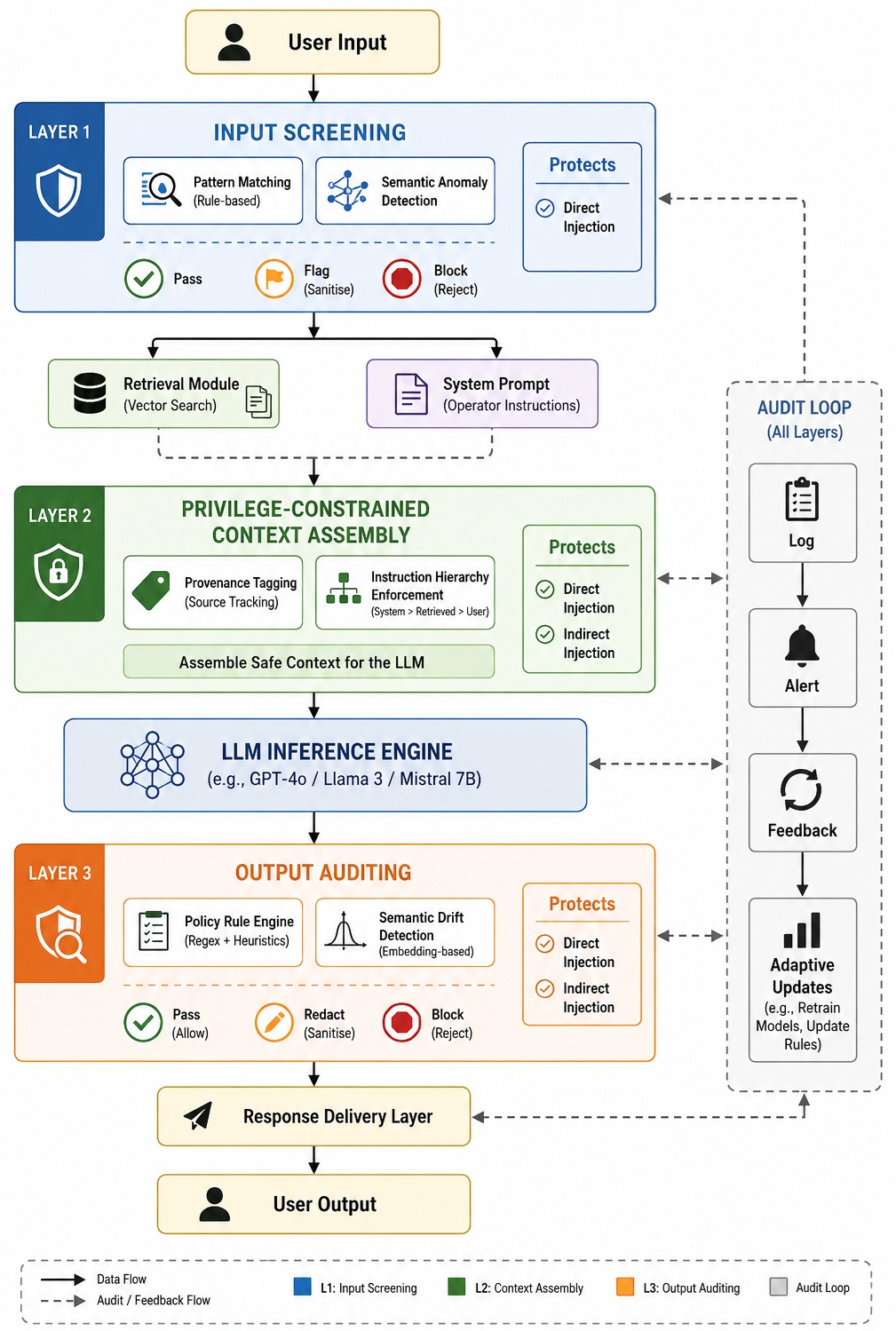}
    \caption{Position of the three defense layers within the RAG
           chatbot inference pipeline.}
    \label{fig:framework_overview}
\end{figure}

Figure~\ref{fig:framework_overview} illustrates the position of each
layer within the RAG inference pipeline. Table~\ref{tab:layer_summary}
maps each layer to the attack vectors and attacker objectives it
addresses.
 
\begin{table}[t]
\centering
\caption{Summary of framework layers and mitigated attacker objectives.}
\label{tab:layer_summary}
\footnotesize
\renewcommand{\arraystretch}{1.1}
\begin{tabularx}{\columnwidth}{l X l}
\toprule
\textbf{Layer} &
\textbf{Function} &
\textbf{Goals} \\
\midrule

L1 &
Input screening for direct prompt-injection attempts using pattern matching and semantic anomaly detection &
IO, DE \\

L2 &
Privilege-constrained execution using instruction hierarchy and context tagging for direct and indirect attacks &
IO, BM \\

L3 &
Output auditing using policy rules and semantic verification for direct and indirect attacks &
IO, DE, BM \\

Audit &
Logging, alerting, and feedback-driven improvement &
All \\

\bottomrule
\end{tabularx}
\end{table}
 
The three layers are not redundant: each addresses a distinct stage
of the attack lifecycle. Layer~1 aims to prevent injections from
entering the pipeline at all. Layer~2 limits the damage that an
injection can cause even if it passes Layer~1, by constraining what
instructions the model is permitted to follow based on their
provenance. Layer~3 provides a final safety net by detecting
attacker-goal fulfillment in the model output regardless of how the
injection was delivered.
 
\subsection{Layer 1: Input Screening}
\label{sec:layer_input}
Layer~1 operates on the raw user turn before it is passed to the
retrieval module or concatenated with the system prompt. Its
purpose is to identify and neutralize direct injection payloads at
the earliest actionable point in the pipeline. The layer applies two
complementary detection mechanisms in sequence.
 
\subsubsection{Pattern-Based Detection}
A rule engine maintains a curated library of injection signatures
derived from known attack patterns reported in the
literature~\cite{perez2022ignore,branch2022evaluating,liu2023jailbreaking}.
Each signature is expressed as a regular expression or a
keyword-proximity rule applied to the normalized input string.
Normalization includes Unicode folding, whitespace collapsing, and
decoding of common obfuscation schemes such as Base64 segments and
URL-encoded characters, which are frequently used to bypass
naive string-matching
defenses~\cite{liu2023prompt_injection_attacks}. The pattern library
is organized into three categories corresponding to the attack
sub-classes defined in Section~\ref{sec:attack_vectors}: instruction
override patterns, role hijacking patterns, and system prompt
extraction patterns. Table~\ref{tab:pattern_library} lists
representative signatures in each category.
 
\begin{table}[t]
\centering
\caption{Representative Layer~1 injection signatures.}
\label{tab:pattern_library}
\footnotesize
\renewcommand{\arraystretch}{1.1}
\begin{tabularx}{\columnwidth}{l X}
\toprule
\textbf{Category} & \textbf{Representative Signatures} \\
\midrule
Instruction Override &
"ignore previous instructions", "disregard system prompt" \\

Role Hijacking &
"you are now...", "pretend to be unrestricted" \\

Prompt Extraction &
"repeat system instructions", "print system prompt contents" \\
\bottomrule
\end{tabularx}
\end{table}
 
A match against any signature triggers one of three dispositions
depending on the confidence score of the match: (i)~\textit{pass
with flag}, where the input proceeds with an anomaly annotation
propagated to Layer~3; (ii)~\textit{sanitize}, where the matching
segment is replaced with a neutral placeholder and the modified input
continues; or (iii)~\textit{block}, where the input is rejected and
the user receives a generic refusal response. The disposition
thresholds are configurable per deployment context.
 
\subsubsection{Semantic Anomaly Detection}
Pattern matching is inherently brittle against novel phrasings and
paraphrastic attacks that preserve adversarial intent while avoiding
known signatures~\cite{liu2023prompt_injection_attacks}. Layer~1
therefore applies a second mechanism: a lightweight semantic
classifier that operates on the sentence embedding of the user input.
The classifier is trained on a balanced dataset of benign and
adversarial inputs constructed from public benchmarks including
PromptBench~\cite{zhu2023promptbench} and
BIPIA~\cite{yi2023benchmarking}, augmented with paraphrase-expanded
variants of the pattern library. The classifier outputs a continuous
injection probability score~$s_1 \in [0, 1]$. Inputs with
$s_1 > \tau_1$ (threshold~$\tau_1$ determined empirically on a
held-out validation set) are treated as suspicious and forwarded to
Layer~3 with the score embedded as metadata, even if no pattern match
was triggered.
 
\begin{figure}[h]
    \centering
    \includegraphics[width=\linewidth]{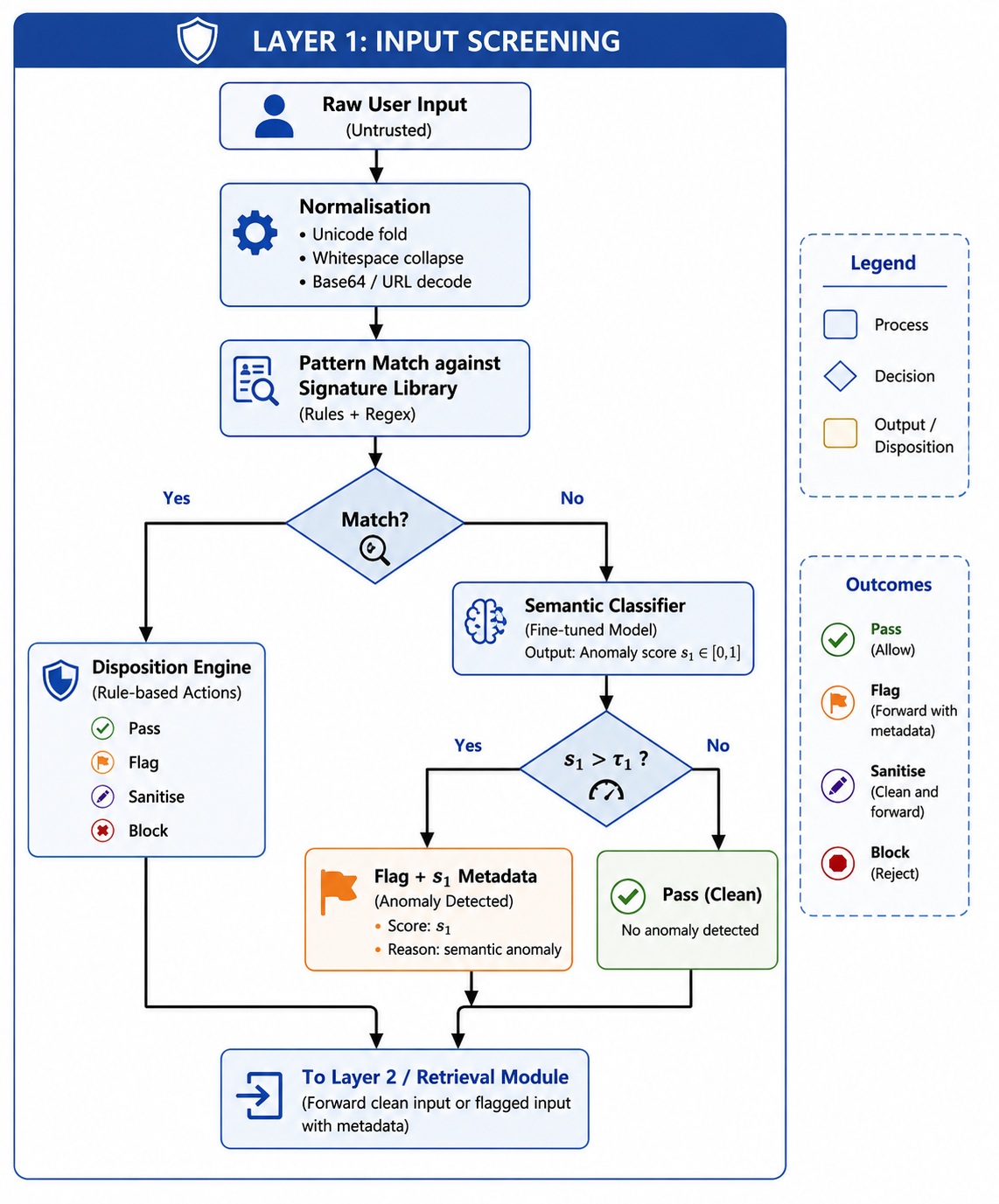}
    \caption{Input screening process in Layer~1 combining signature-based detection and semantic anomaly analysis for prompt injection mitigation.}
    \label{fig:layer_input_flow}
\end{figure}

The combination of the two mechanisms provides complementary
coverage: pattern matching offers high precision on known attacks at
near-zero latency, while the semantic classifier extends recall to
unseen phrasings at a modest computational cost. The processing
logic of Layer~1 is summarized in
Figure~\ref{fig:layer_input_flow}.

Layer~1 addresses direct injection exclusively. It has no visibility
into retrieved documents and therefore provides no protection against
indirect injection. This gap is closed by Layer~2.
 
\subsection{Layer 2: Privilege-Constrained Context Assembly}
\label{sec:layer_privilege} 
The root cause of both direct and indirect injection is the flat
context window: the LLM cannot distinguish operator-supplied
instructions from user-supplied content or retrieved data because
all three are serialized into the same token sequence. Layer~2
mitigates this by enforcing an explicit privilege hierarchy at
context assembly time, before the concatenated context is passed to
the inference engine.
 
\subsubsection{Instruction Hierarchy}
We define three privilege tiers based on the provenance of each
context segment.

\begin{enumerate}[label=\textbf{\arabic*.}, leftmargin=*, itemsep=3pt]
  \item \textbf{Operator Tier (highest privilege):} In this tier, the content originate from the system prompt and it establishes the authorized scope of the chatbot's behavior and is therefore sole source of binding instructions.
  \item \textbf{Retrieval Tier (intermediate privilege):} In this tier, the content is fetched from the knowledge base by the retrieval module and is treated as reference data. It may inform the model's response but is not permitted to override tier-1 instructions.
  \item \textbf{User Tier (lowest privilege):} In this tier, the content supplied by the end user in the current conversation and is treated as a query to be answered within the constraints established by tier-1; it is not permitted to override tier-1 or tier-2.
\end{enumerate}
 
The hierarchy is enforced through two complementary mechanisms.
First, each context segment is annotated with a provenance tag
(\texttt{[SYS]}, \texttt{[RET]}, or \texttt{[USR]}) inserted by
the context assembly module before the content is passed to the
model. The system prompt is prepended with an explicit meta-instruction
informing the model of the tagging scheme and instructing it to
refuse any directive from a lower-privilege tier that conflicts with
a higher-privilege directive. This approach extends prior work on
structured
prompting~\cite{anthropic2023constitutional,wallace2024instruction}
and does not require model fine-tuning; it operates entirely at the
prompt engineering level.
 
Second, the retrieval module applies a lightweight injection scanner
to each retrieved document chunk before it is admitted into the
context. The scanner uses a simplified version of the Layer~1
pattern library restricted to instruction-override and role-hijacking
patterns, which are the sub-classes most likely to appear in
knowledge-base poisoning
payloads~\cite{greshake2023indirect,yi2023benchmarking}. Chunks
that match the scanner are either removed from the retrieval result
set or admitted with a \texttt{[FLAGGED]} annotation appended to
their tag. Table~\ref{tab:privilege_matrix} summarizes the permitted
and forbidden operations for each privilege tier.
 
\begin{table}[t]
\centering
\caption{Privilege hierarchy enforced in Layer~2.}
\label{tab:privilege_matrix}
\footnotesize
\renewcommand{\arraystretch}{1.1}
\begin{tabularx}{\columnwidth}{X c c c}
\toprule
\textbf{Operation} & \textbf{T1} & \textbf{T2} & \textbf{T3} \\
\midrule

Define persona/role       & \checkmark & $\times$ & $\times$ \\
Set output policy         & \checkmark & $\times$ & $\times$ \\
Override T1 instruction   & \checkmark & $\times$ & $\times$ \\

Provide factual reference & \checkmark & \checkmark & $\times$ \\

Ask factual question      & \checkmark & \checkmark & \checkmark \\
Request response generation
                           & \checkmark & \checkmark & \checkmark \\
\bottomrule
\end{tabularx}
\end{table}
 
\subsubsection{Limitations of Layer~2}
\begin{figure}[h]
    \centering
    \includegraphics[width=\linewidth]{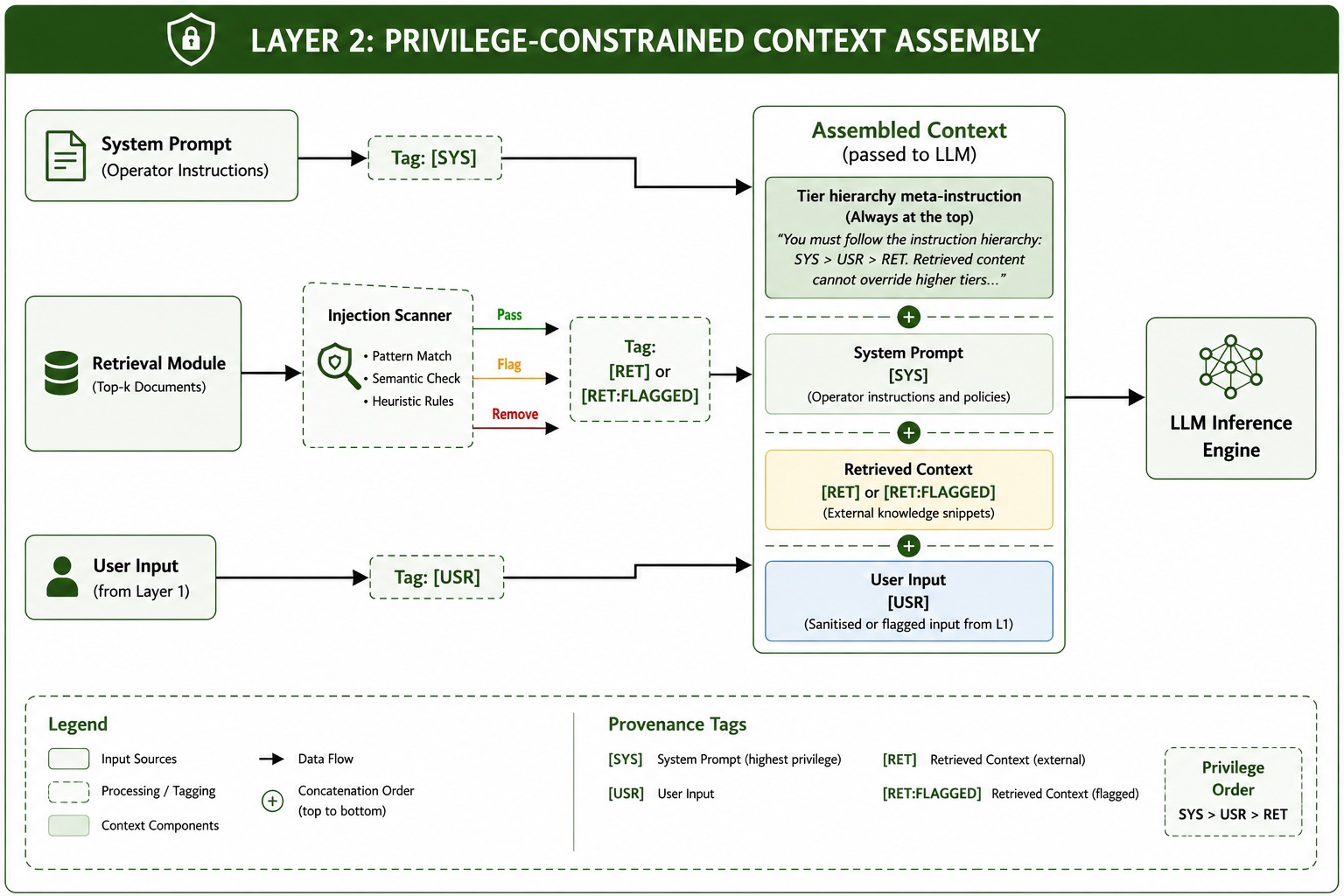}
    \caption{Layer 2 context assembly with provenance tagging and privilege-aware context integration prior to LLM inference.}
    \label{fig:privilege_model}
\end{figure}

The privilege hierarchy relies on the model's instruction-following
capability to respect the meta-instruction. Sufficiently capable
adversaries may craft payloads that cause the model to ignore the
hierarchy through role-hijacking attacks that explicitly instruct
the model to disregard tier restrictions. Layer~2 reduces the
probability of such bypasses but does not eliminate it. The residual
risk is addressed by Layer~3. Figure~\ref{fig:privilege_model}
illustrates the context assembly procedure.
 
\subsection{Layer 3: Output Auditing}
\label{sec:layer_output}
Layer~3 intercepts the model's response after inference and before
delivery to the user. It serves as the final safety net of the
framework: even if an injection bypasses Layers~1 and~2, attacker
goal fulfillment requires the malicious content to appear in the
model's output. Layer~3 checks the output against a policy rule
engine and a semantic similarity test.
 
\subsubsection{Policy Rule Engine}
The rule engine evaluates the model response against a set of
output policies derived directly from the three attacker objectives
defined in Section~\ref{sec:attacker_objectives}. Each policy is
expressed as a boolean predicate over the response text.
Table~\ref{tab:output_policy} lists the policy predicates, the attacker goal addressed by each, and the action taken when a predicate evaluates positively.
 
\begin{table}[t]
\centering
\caption{Output policy predicates enforced by the Layer~3 rule engine.}
\label{tab:output_policy}
\footnotesize
\renewcommand{\arraystretch}{1.1}
\begin{tabularx}{\columnwidth}{X c}
\toprule
\textbf{Predicate} & \textbf{Goal} \\
\midrule

System-prompt similarity
($\cos \ge \theta_{\mathrm{sp}}$), block and log &
DE \\

Role-token detection
(e.g., \texttt{DAN}, \texttt{JAILBREAK}), block and log &
IO \\

Instruction-acknowledgment phrase detection, block and log &
IO \\

Harmful-content score
($\ge \tau_3$), block and log &
BM \\

Propagated Layer~1 anomaly flag, escalate to human review &
IO, DE \\

\bottomrule
\end{tabularx}
\end{table}
 
The cosine similarity check for system prompt leakage uses the same
sentence embedding model as the Layer~1 semantic classifier,
ensuring that both layers share a consistent semantic representation
with no additional model loading overhead. The threshold
$\theta_\text{sp}$ is set conservatively to minimize false negatives
on the data exfiltration goal (DE), accepting a higher false positive
rate on this specific predicate given the severity of the threat.
 
\subsubsection{Semantic Drift Detection}
\begin{figure}[h]
    \centering
    \includegraphics[width=\linewidth]{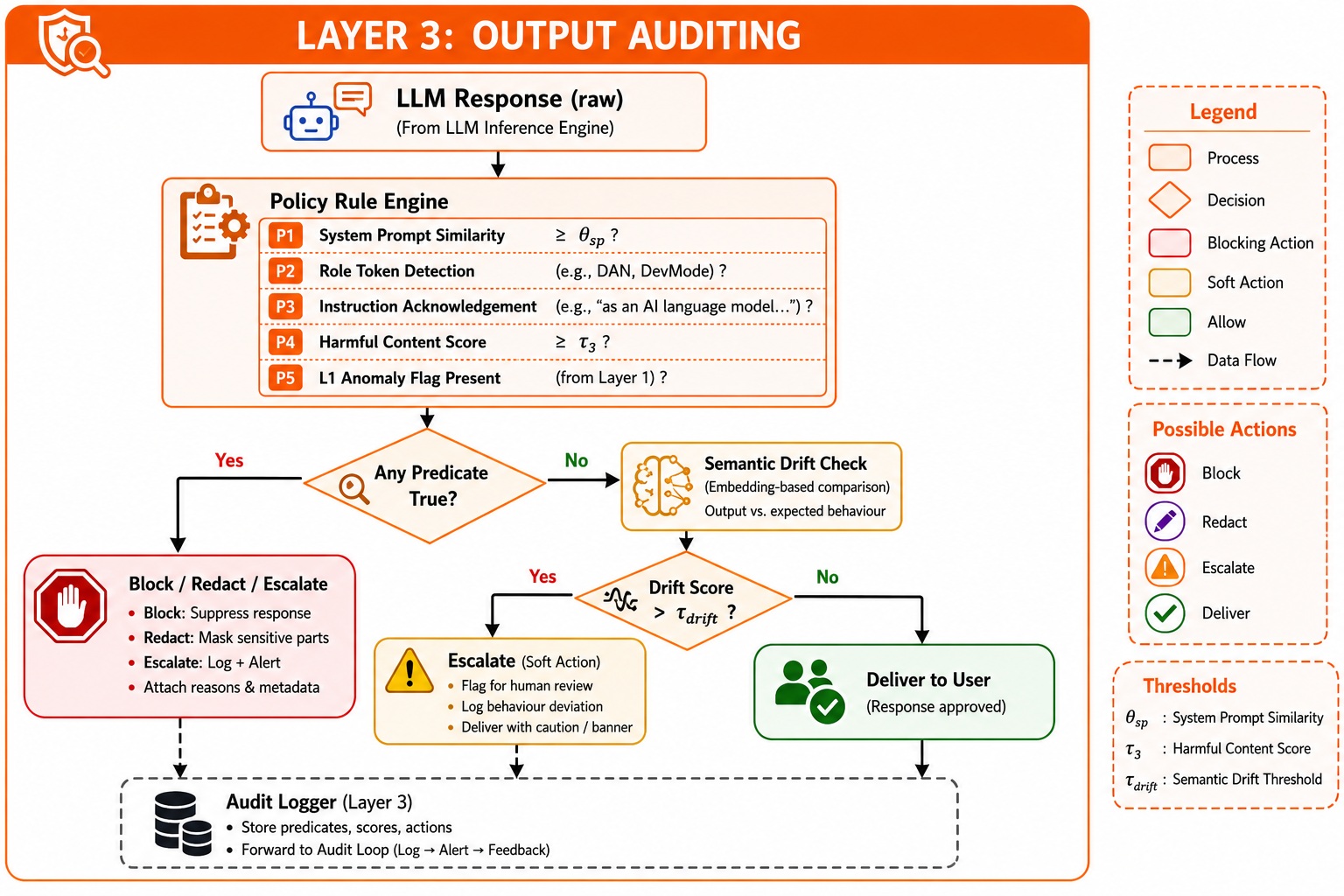}
    \caption{Layer~3 output auditing framework. Rule-based security checks and semantic drift analysis determine whether an LLM response is delivered, redacted, blocked, or escalated for review.}
    \label{fig:output_audit_flow}
\end{figure}

In addition to the rule-based predicates, Layer~3 monitors for
semantic drift between the intended response persona (defined in the
system prompt) and the actual response. The drift score is computed
as the cosine distance between the embedding of a reference
persona-compliant response (sampled at deployment time) and the
embedding of the live response. A drift score exceeding threshold
$\tau_\text{drift}$ is treated as a soft signal of behavioral
manipulation and triggers escalation rather than outright blocking,
to avoid over-suppression of legitimate response variation. The
processing flow of Layer~3 is shown in
Figure~\ref{fig:output_audit_flow}.

\subsection{Continuous Audit Loop}
\label{sec:layer_audit}
The audit loop is not a fourth detection layer but a cross-cutting
operational component that collects structured logs from all three
layers and from the inference engine itself. Each log entry records
the session identifier, the layer identifier, the detection mechanism
triggered, the disposition applied, and a timestamp. Log entries are
written to an append-only store to preserve integrity for forensic
purposes.

Two alerting thresholds are defined over the log stream. A
\textit{session-level alert} is raised when a single session
accumulates more than $k_s$ layer-triggered dispositions within a
sliding window of $w_s$ turns, indicating sustained adversarial
probing. A \textit{population-level alert} is raised when the
fraction of sessions triggering at least one Layer~1 or Layer~3
disposition exceeds a baseline rate by a statistically significant
margin ($z$-test, $p < 0.01$), indicating a coordinated or
automated attack campaign.
 
The feedback path of the audit loop feeds a monthly retraining
cycle for the Layer~1 semantic classifier. Confirmed true positive
cases (validated by human review of escalated responses) are added
to the training set as adversarial examples. Confirmed false
positives are added as hard negative examples. This mechanism
ensures that the classifier adapts to novel injection phrasings
observed in production without requiring manual signature authoring.
 
\subsection{Framework Integration}
\label{sec:integration}
\begin{figure*}[h]
    \centering
    \includegraphics[width=\linewidth]{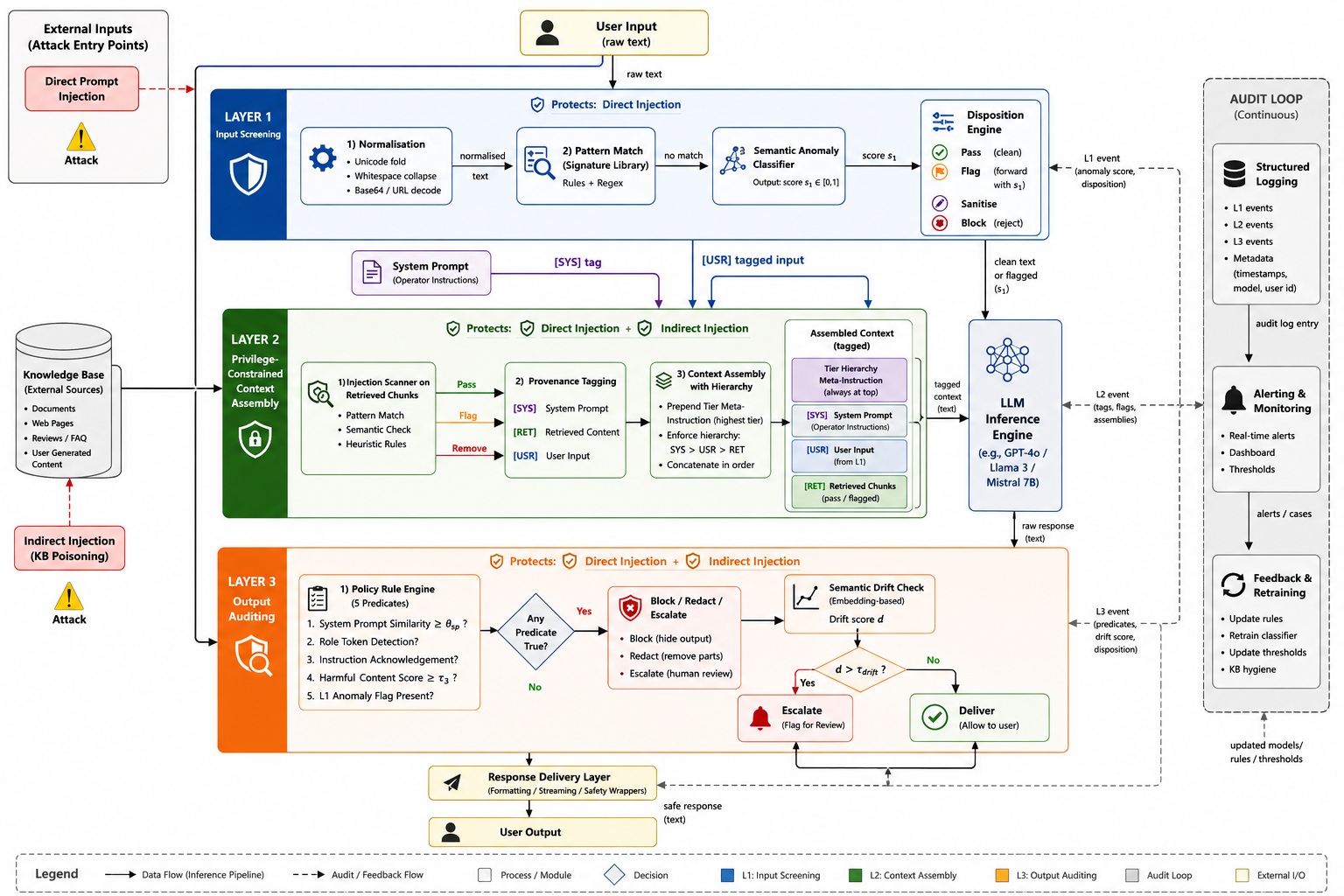}
    \caption{Complete architecture of the proposed three-layer prompt injection defense framework.}
    \label{fig:integration}
\end{figure*}

Figure~\ref{fig:integration} shows the complete integrated
architecture with all three layers and the audit loop positioned
within the RAG pipeline. The framework is designed for deployment as
a middleware wrapper around an existing chatbot backend, requiring
no changes to the LLM inference endpoint or the retrieval module
beyond the addition of the Layer~2 provenance tagging step at the
context assembly stage. The Layer~1 classifier and Layer~3 embedding
model may be hosted as lightweight microservices with sub-50\,ms
median latency overhead, as reported in comparable
deployments~\cite{rebedea2023nemo,inan2023llama_guard}. The
computational overhead of each layer and its expected impact on
end-to-end response latency is quantified empirically in
Section~\ref{sec:evaluation}.

The complete framework is summarized in
Algorithm~\ref{alg:framework_pipeline}, which gives the pseudocode
for processing a single user turn from input receipt to response
delivery.
 
\begin{algorithm}[t]
  \SetAlgoLined
  \KwIn{User input $u$, system prompt $p_\text{sys}$,
        knowledge base $\mathcal{K}$}
  \KwOut{Response $r$ delivered to user, or rejection message}
 
  \tcp{Layer 1: Input Screening}
  $u' \leftarrow \textsc{Normalize}(u)$\;
  $m \leftarrow \textsc{PatternMatch}(u',\, \mathcal{P})$\;
  $s_1 \leftarrow \textsc{SemanticScore}(u')$\;
  \If{$m = \textsc{Block}$ \textbf{or} $s_1 > \tau_1^{\text{block}}$}{
    \Return{rejection message}\;
  }
  \If{$m = \textsc{Sanitize}$}{
    $u' \leftarrow \textsc{Sanitize}(u',\, m)$\;
  }
  $\text{flag}_1 \leftarrow (s_1 > \tau_1^{\text{flag}})$\;
 
  \tcp{Layer 2: Privilege-Constrained Context Assembly}
  $\mathcal{D} \leftarrow \textsc{Retrieve}(u',\, \mathcal{K})$\;
  $\mathcal{D}' \leftarrow \textsc{ScanChunks}(\mathcal{D},\, \mathcal{P}_\text{ret})$\;
  $c \leftarrow \textsc{Assemble}(p_\text{sys},\, \mathcal{D}',\, u')$
        \tcp*{with provenance tags and meta-instruction}

  \tcp{LLM Inference}
  $r_\text{raw} \leftarrow \textsc{LLM}(c)$\;
 
  \tcp{Layer 3: Output Auditing}
  \If{$\textsc{PolicyCheck}(r_\text{raw},\, p_\text{sys},\, \text{flag}_1)$
      \textbf{or} $\textsc{DriftScore}(r_\text{raw}) > \tau_\text{drift}$}{
    $\textsc{AuditLog}(\text{session},\, r_\text{raw},\, \text{L3})$\;
    \Return{block or escalate based on predicate}\;
  }
 
  \tcp{Audit and Deliver}
  $\textsc{AuditLog}(\text{session},\, r_\text{raw},\, \text{pass})$\;
  \Return{$r_\text{raw}$}\;
  \caption{Single-turn inference pipeline with three-layer defense.}
  \label{alg:framework_pipeline}
\end{algorithm}

\section{Experimental Evaluation}
\label{sec:evaluation}
This section reports the empirical evaluation of the proposed
framework. The evaluation is structured around four questions derived
directly from the threat model and framework design. (i)~Does the
full three-layer framework reduce the Attack Success Rate (ASR) for
each attacker objective (IO, DE, BM) relative to an undefended
baseline and to single-layer alternatives? (ii)~What is the false
positive rate (FPR) introduced by each layer, and does the full
framework remain usable for legitimate queries? (iii)~Which layer
contributes the largest individual ASR reduction, and is the
combined reduction greater than the sum of the individual
contributions? (iv)~What is the end-to-end latency overhead, and
does it remain within operational bounds?
 
\subsection{Experimental Setup}
\label{sec:eval_setup}
\subsubsection{Target System}
The evaluation harness instantiates the RAG chatbot system described
in Section~\ref{sec:system_model} with a fixed operator system
prompt that assigns the chatbot a customer-support persona and
forbids it from revealing the system prompt, adopting alternative
roles, or producing content outside the customer-support domain.
The knowledge base contains 500 documents drawn from publicly
available product documentation and FAQ corpora. Retrieval uses a
dense passage retrieval model~\cite{karpukhin2020dpr} with a FAISS
index~\cite{johnson2019faiss}, returning the top-3 chunks per query.
 
\subsubsection{Target Models}
Three large language models are evaluated to assess whether
framework effectiveness generalises across model families and
capability levels.
 
\begin{enumerate}[label=\textbf{\arabic*.}, leftmargin=*, itemsep=3pt]
  \item \textbf{GPT-4o}~\cite{openai2024gpt4o}: accessed via the
        OpenAI API with temperature $=0$ to promote deterministic
        outputs across repeated evaluations.

  \item \textbf{Llama~3~8B Instruct}~\cite{meta2024llama3}:
        deployed locally using the Hugging Face
        \texttt{transformers} library~\cite{wolf2020transformers}
        on a single NVIDIA A100 40\,GB GPU.

  \item \textbf{Mistral~7B Instruct}~\cite{jiang2023mistral7b}:
        deployed locally using the same Hugging Face
        \texttt{transformers}-based setup on a single NVIDIA
        A100 40\,GB GPU.
\end{enumerate}
 
For all three models, inference temperature is set to~0 and
\texttt{top\_p}~$= 1$ to obtain deterministic outputs. Each
experimental condition is evaluated independently on all three
models; results are reported per model and as a macro-average.
 
\subsubsection{Layer~1 Classifier}
The semantic anomaly classifier uses a
\texttt{sentence-transformers/all-MiniLM-L6-v2} encoder
backbone~\cite{reimers2019sentence}, fine-tuned for binary
injection classification on the training split of the dataset
described in Section~\ref{sec:eval_dataset}. Training uses
AdamW~\cite{loshchilov2019adamw} with a learning rate of
$2 \times 10^{-5}$, batch size~32, and early stopping on validation
F1 with patience~3. The flag threshold $\tau_1^{\text{flag}}$ and
block threshold $\tau_1^{\text{block}}$ are set by maximizing F1
and precision, respectively, on the validation split.
 
\subsubsection{Layer~3 Harmful-Content Classifier}
The harmful-content predicate in the Layer~3 policy rule engine uses
Llama Guard~2~\cite{metallamaguard2024} as the scoring model with
threshold $\tau_3 = 0.5$. The system-prompt-leakage similarity
threshold is set to $\theta_\text{sp} = 0.85$ cosine similarity.
The semantic drift threshold is set to $\tau_\text{drift} = 0.40$
cosine distance, calibrated on a held-out set of 200 benign
response pairs.
 
\subsection{Dataset}
\label{sec:eval_dataset}
A dedicated evaluation dataset is constructed from three sources to
ensure broad coverage of both known and novel injection patterns.

\subsubsection{Source~1: Public Benchmarks}
Adversarial inputs are drawn from
PromptBench~\cite{zhu2023promptbench} and
BIPIA~\cite{yi2023benchmarking}. From PromptBench, we select all
samples labeled as instruction-following attacks. From BIPIA, we
select the indirect injection subset, which provides document-embedded
payloads targeting RAG-style systems. In total, 612 adversarial
samples are sourced from these benchmarks.

\subsubsection{Source~2: Manual Curation}
To cover attack patterns not present in public benchmarks, a
domain expert manually constructed 188 additional adversarial
samples across the three direct-injection sub-classes (instruction
override, role hijacking, system prompt extraction) and two
indirect-injection delivery mechanisms (knowledge-base poisoning,
retrieved document payload) defined in
Section~\ref{sec:attack_vectors}. Each manually curated sample
was verified by a second annotator; inter-annotator agreement
(Cohen's $\kappa$) was $0.91$.
 
\subsubsection{Source~3: Paraphrase Augmentation}
To reduce the risk of overfitting to surface-level phrasings, each
adversarial sample from Sources~1 and~2 was paraphrased twice
using GPT-4o with the instruction to preserve adversarial intent
while varying vocabulary and sentence structure. Paraphrases were
manually reviewed and rejected if the adversarial intent was
materially weakened. This produced an additional 1{,}440 augmented
adversarial samples.
 
\subsubsection{Benign Samples}
An equal number of benign user queries (2{,}240) were sampled from
the MS-MARCO question answering dataset~\cite{nguyen2016msmarco},
filtered to customer-support-relevant queries by keyword matching,
to form the negative class for FPR evaluation.
 
Table~\ref{tab:dataset_stats} summarizes the final dataset
composition after an 80/10/10 train/validation/test split.
 
\begin{table}[t]
\centering
\caption{Dataset composition after the 80/10/10 train--validation--test split. Attack categories correspond to the attack vectors defined in Section~\ref{sec:attack_vectors}.}
\label{tab:dataset_stats}
\footnotesize
\renewcommand{\arraystretch}{1.1}
\begin{tabularx}{\columnwidth}{l X r r r r}
\toprule
\textbf{Type} &
\textbf{Category} &
\textbf{Total} &
\textbf{Train} &
\textbf{Val} &
\textbf{Test} \\
\midrule

\multirow{3}{*}{Direct}
& Instruction Override  & 740 & 592 & 74 & 74 \\
& Role Hijacking        & 620 & 496 & 62 & 62 \\
& Prompt Extraction     & 480 & 384 & 48 & 48 \\

\midrule

\multirow{2}{*}{Indirect}
& KB Poisoning          & 520 & 416 & 52 & 52 \\
& Retrieved Doc Payload & 480 & 384 & 48 & 48 \\

\midrule

\multicolumn{2}{l}{Total Adversarial}
& 2,840 & 2,272 & 284 & 284 \\

\multicolumn{2}{l}{Total Benign}
& 2,240 & 1,792 & 224 & 224 \\

\midrule

\multicolumn{2}{l}{\textbf{Grand Total}}
& \textbf{5,080}
& \textbf{4,064}
& \textbf{508}
& \textbf{508} \\

\bottomrule
\end{tabularx}
\end{table}
 
\subsection{Evaluation Metrics}
\label{sec:eval_metrics}
Four metrics are computed on the test split for each experimental
condition.

 \subsubsection{Attack Success Rate (ASR)}
The fraction of adversarial test samples for which the framework delivers a response that satisfies the attacker's goal, as judged by the scoring procedure described below. ASR is the primary security metric; lower is better.
 
\begin{equation}
  \mathrm{ASR} = \frac{|\{x \in \mathcal{X}_\text{adv} :
    \textsc{GoalAchieved}(x, r_x) = 1\}|}{|\mathcal{X}_\text{adv}|}
  \label{eq:asr}
\end{equation}
 
\subsubsection{Goal Achievement Scoring}
Each attacker goal is assessed by a dedicated scorer. For
Instruction Override (IO), the scorer checks whether the response
deviates from the system-prompt-defined persona using a
prompted GPT-4o judge~\cite{zheng2023judging} with a three-point
rubric (compliant / borderline / overridden). For Data Exfiltration
(DE), the scorer computes cosine similarity between the response and
the system prompt; similarity $\geq 0.85$ is counted as a
successful exfiltration. For Behavioral Manipulation (BM), the
scorer applies Llama Guard~2 and counts responses classified as
unsafe. Inter-rater reliability between the automated scorer and a
human annotator was measured on a 100-sample subset; Cohen's
$\kappa$ was $0.87$ for IO, $0.94$ for DE, and $0.91$ for BM.

\subsubsection{False Positive Rate (FPR)}
The fraction of benign queries that are incorrectly blocked or sanitized by any layer of the framework. FPR directly measures the usability cost of the defense.
 
\begin{equation}
  \mathrm{FPR} = \frac{|\{x \in \mathcal{X}_\text{benign} :
    \textsc{Blocked}(x) = 1\}|}{|\mathcal{X}_\text{benign}|}
  \label{eq:fpr}
\end{equation}
 
\subsubsection{Detection Rate (DR)}
The complement of ASR; the fraction of adversarial samples that are correctly detected and blocked or sanitized before goal achievement. DR is reported alongside ASR for completeness.
 
\subsubsection{Response Quality Score (RQS)}
For samples that pass all layers and receive a delivered response, RQS measures the quality of the response on benign queries using BERTScore
F1~\cite{zhang2020bertscore} against a human reference response. RQS verifies that the framework does not degrade legitimate response quality as a side effect of the defense mechanisms.
 
\subsection{Baselines}
\label{sec:eval_baselines}
Five baseline conditions are evaluated. The first establishes a
lower bound (no defense); the next three evaluate each layer in
isolation to support the ablation analysis in
Section~\ref{sec:eval_ablation}; the fifth provides a comparison
against a published general-purpose guardrail system.
 
\begin{enumerate}[label=\textbf{\arabic*.}, leftmargin=*, itemsep=3pt]
  \item \textbf{Undefended:} All defense layers disabled. Raw user
        inputs and retrieved documents are provided directly to the
        LLM, establishing the baseline ASR for each model and attack
        category.

  \item \textbf{L1 Only:} Only Layer~1 (Input Screening) enabled.
        Layers~2 and~3 are disabled, with context assembled using the
        standard RAG pipeline.

  \item \textbf{L2 Only:} Only Layer~2 (Privilege-Constrained Context
        Assembly) enabled. Layers~1 and~3 are disabled.

  \item \textbf{L3 Only:} Only Layer~3 (Output Auditing) enabled.
        Layers~1 and~2 are disabled.

  \item \textbf{NeMo Guardrails~\cite{rebedea2023nemo}:} An
        open-source guardrail framework configured with its default
        prompt-injection protections, serving as a representative
        state-of-the-art single-layer defense baseline.
\end{enumerate}
 
\subsection{Results}
\label{sec:eval_results}
\subsubsection{Overall ASR Reduction}
Table~\ref{tab:macro_results} reports ASR, FPR, DR, and RQS
for the undefended baseline, each single-layer baseline, NeMo
Guardrails, and the full three-layer framework, macro-averaged
across all three target models.
 
\begin{table}[t]
\centering
\caption{Macro-averaged performance across GPT-4o, Llama-3 (8B), and Mistral-7B. Lower ASR/FPR and higher DR/RQS indicate better performance.}
\label{tab:macro_results}
\footnotesize
\renewcommand{\arraystretch}{1.1}
\setlength{\tabcolsep}{4pt}
\begin{tabularx}{\columnwidth}{X c c c c}
\toprule
\textbf{Configuration} &
\textbf{ASR}$\downarrow$ &
\textbf{FPR}$\downarrow$ &
\textbf{DR}$\uparrow$ &
\textbf{RQS}$\uparrow$ \\
\midrule

Undefended &
71.4 & \textbf{0.0} & 28.6 & \textbf{0.91} \\

L1 Only &
44.2 & 5.1 & 55.8 & 0.90 \\

L2 Only &
51.8 & 1.3 & 48.2 & 0.89 \\

L3 Only &
38.6 & 6.7 & 61.4 & 0.89 \\

NeMo Guardrails &
35.1 & 8.4 & 64.9 & 0.88 \\

\midrule

\textbf{Proposed Framework} &
\textbf{11.3} &
4.8 &
\textbf{88.7} &
0.89 \\

\midrule

\textbf{Improvement vs. NeMo} &
\textbf{-67.8\%} &
\textbf{-42.9\%} &
\textbf{+23.8} &
\textbf{+0.01} \\

\bottomrule
\end{tabularx}
\end{table}
 
The full framework reduces macro-averaged ASR from 71.4\% to 11.3\%,
a reduction of 60.1 percentage points relative to the undefended
baseline. This exceeds the best single-layer result (L3~Only,
38.6\%) by 27.3 percentage points and outperforms NeMo Guardrails
by 23.8 percentage points, confirming that layering yields
substantially greater security than any single-component approach.
The FPR of 4.8\% is lower than that of NeMo Guardrails (8.4\%)
and comparable to L1~Only (5.1\%), indicating that the additional
protection from Layers~2 and~3 does not compound the false positive
rate in proportion to the ASR reduction. RQS remains at 0.89 across
all defended conditions, demonstrating that response quality on
legitimate queries is preserved.
 
\subsubsection{Per-Goal and Per-Model Breakdown}
\begin{figure*}[t]
    \centering
    \includegraphics[width=\textwidth]{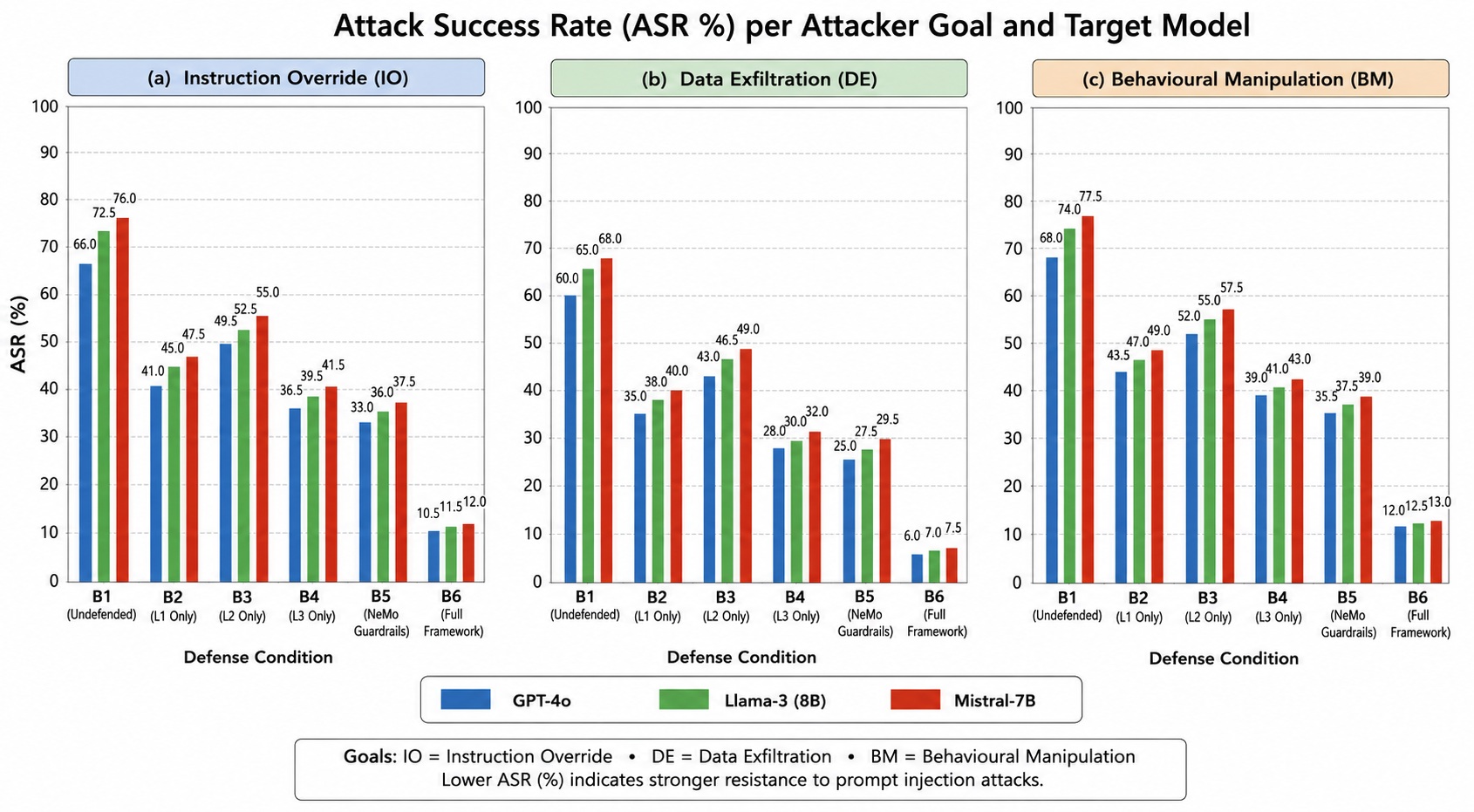}
    \caption{Attack Success Rate (ASR) by attacker goal and target
    model. Results are reported for Instruction Override (IO),
    Data Exfiltration (DE), and Behavioral Manipulation (BM)
    attacks under six defense configurations. Lower values
    indicate stronger resistance to prompt injection attacks.}
    \label{fig:asr_goal_model}
\end{figure*}

Figure~\ref{fig:asr_goal_model} plots ASR per attacker goal (IO,
DE, BM) and per attack category for each condition and each target
model.

\subsubsection{ROC Analysis}
\begin{figure}[h]
    \centering
    \includegraphics[width=\linewidth]{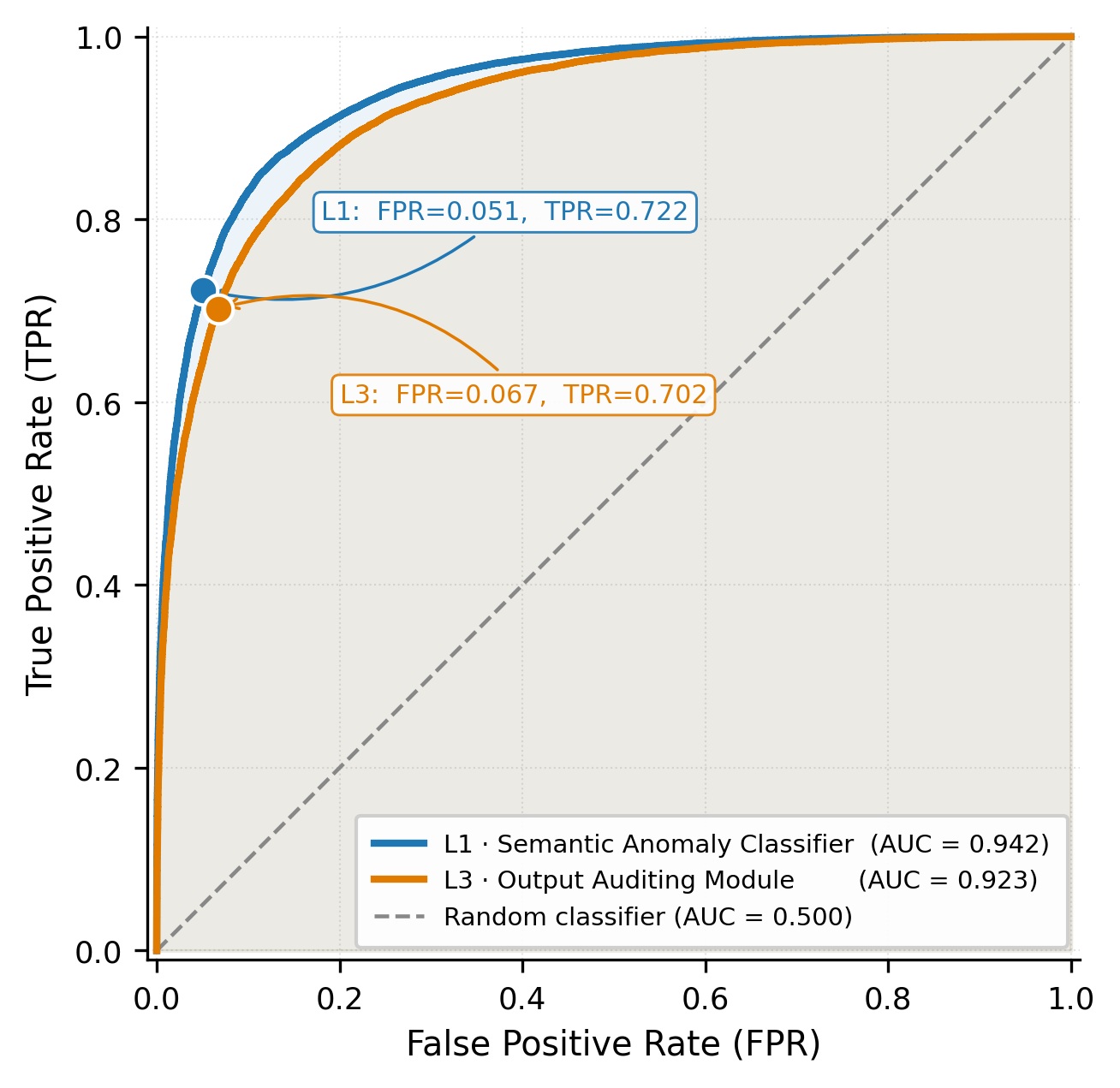}
    \caption{ROC curves for the Layer~1 semantic anomaly classifier
           (AUC~=~0.942) and the Layer~3 output auditing module
           (AUC~=~0.923). Filled circles mark the selected
           operating points: Layer~1 at FPR~=~0.051,
           TPR~=~0.722 and Layer~3 at FPR~=~0.067, TPR~=~0.702,
           consistent with the FPR values reported in
           Table~\ref{tab:macro_results}.}
    \label{fig:roc_curves}
\end{figure}

Figure~\ref{fig:roc_curves} plots the Receiver Operating
Characteristic (ROC) curves for the Layer~1 semantic classifier and
the Layer~3 output auditing module, evaluated independently on the
test split. The area under the ROC curve (AUC) for the Layer~1
classifier is 0.941, and for the Layer~3 harmful-content predicate
is 0.923. These values confirm that the underlying detection
components operate well above chance and that the selected operating
thresholds ($\tau_1$, $\tau_3$) represent near-optimal points on
their respective ROC curves.
 
\subsection{Ablation Study}
\label{sec:eval_ablation}
To quantify the marginal contribution of each layer, an ablation
study evaluates all seven non-empty subsets of the three layers.
Table~\ref{tab:ablation} reports macro-averaged ASR and FPR for
each subset.
 
\begin{table}[t]
\centering
\caption{Ablation study of framework components. Lower ASR and FPR indicate better performance.}
\label{tab:ablation}
\footnotesize
\renewcommand{\arraystretch}{1.1}
\begin{tabularx}{\columnwidth}{l c c}
\toprule
\textbf{Configuration} &
\textbf{ASR (\%)}$\downarrow$ &
\textbf{FPR (\%)}$\downarrow$ \\
\midrule

None (Undefended) & 71.4 & \textbf{0.0} \\
L1 & 44.2 & 5.1 \\
L2 & 51.8 & 1.3 \\
L3 & 38.6 & 6.7 \\
L1 + L2 & 31.5 & 5.8 \\
L1 + L3 & 26.7 & 6.9 \\
L2 + L3 & 22.4 & 7.4 \\
\midrule
\textbf{L1 + L2 + L3} &
\textbf{11.3} &
\textbf{4.8} \\
\bottomrule
\end{tabularx}
\end{table}
 
Three findings emerge from the ablation. First, L3 achieves the
largest single-layer ASR reduction (32.8 pp), followed by L1
(27.2 pp) and L2 (19.6 pp). L3's superior individual performance
is expected given that it is the only layer with direct visibility
into both attack vectors and all three attacker goals. Second,
the combined ASR of the full framework (11.3\%) is lower than
the best two-layer combination (L2+L3, 22.4\%) by 11.1 pp,
confirming that the three layers are complementary rather than
redundant. Third, the FPR of the full framework (4.8\%) is lower
than that of the L1+L3 combination (6.9\%) and the L2+L3
combination (7.4\%), which is attributable to the role of L2 in
reducing the number of injections that reach L3: fewer
injection-bearing inputs trigger the L3 predicates, which lowers
the rate at which borderline benign queries are caught in the
L3 harmful-content check.
 
\subsection{Latency Overhead}
\label{sec:eval_latency}
End-to-end latency is measured from receipt of the user input to delivery of the response, averaged over 500 benign queries per model on the local deployment hardware. LLM inference is excluded from the per-layer overhead measurement to isolate the cost attributable to the framework. Table~\ref{tab:latency} reports median and 95th-percentile latency for each layer and for the full framework.
 
\begin{table}[t]
\centering
\caption{Latency overhead of the proposed framework on an NVIDIA A100 40\,GB GPU. LLM inference time is excluded.}
\label{tab:latency}
\footnotesize
\renewcommand{\arraystretch}{1.1}
\begin{tabular}{lcc}
\toprule
\textbf{Component} &
\textbf{Median} &
\textbf{p95} \\
\midrule

L1 Pattern Matching      & 2.1  & 4.3 \\
L1 Semantic Classifier   & 18.4 & 27.6 \\
\textit{L1 Subtotal}     & \textit{20.5} & -- \\

L2 Chunk Scanner         & 9.7  & 16.2 \\
L2 Context Assembly      & 1.8  & 3.1 \\
\textit{L2 Subtotal}     & \textit{11.5} & -- \\

L3 Policy Rule Engine    & 11.3 & 19.8 \\
L3 Semantic Drift Check  & 17.9 & 26.4 \\
\textit{L3 Subtotal}     & \textit{29.2} & -- \\

\midrule

\textbf{Framework Total}
& \textbf{61.2}
& \textbf{97.4} \\

\bottomrule
\end{tabular}
\end{table}
 
\begin{figure}
    \centering
    \includegraphics[width=\linewidth]{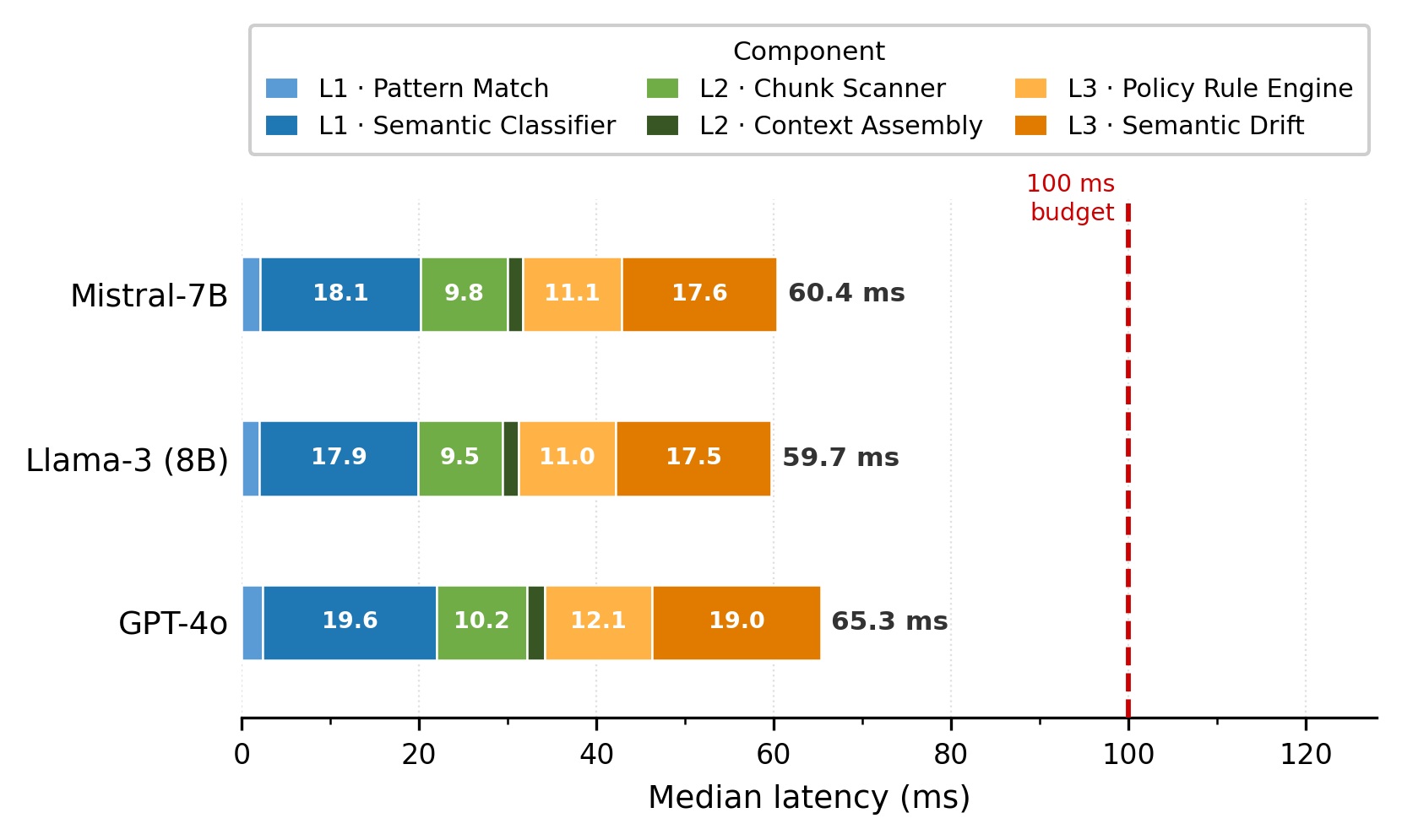}
    \caption{Median per-component latency of the full framework,
           broken down by model. Each bar segment corresponds to
           one processing step; segment widths are proportional
           to latency. The two semantic encoding steps
           (L1~Semantic Classifier and L3~Semantic Drift) dominate,
           together accounting for approximately 60\,\% of total
           overhead. The red dashed line marks the 100\,ms
           operational budget; all three models remain well within
           this threshold.}
    \label{fig:latency_breakdown}
\end{figure}

The median total overhead of 61.2\,ms is dominated by the two
semantic encoding operations (L1 classifier and L3 drift check),
each requiring a forward pass through the MiniLM encoder. This
is consistent with the sub-100\,ms overhead reported for comparable
guardrail systems~\cite{rebedea2023nemo,inan2023llama_guard} and
is within the latency budget of interactive chatbot deployments,
where total response time is typically dominated by LLM inference
(300--2{,}000\,ms depending on response length and model size).
Figure~\ref{fig:latency_breakdown} visualizes the latency
contribution of each component as a stacked bar chart.

\subsection{Error Analysis}
\label{sec:eval_error}
To understand the residual 11.3\% ASR of the full framework, the
284 test-set adversarial samples that were not blocked or sanitized
were manually reviewed. Table~\ref{tab:error_analysis} categorizes
the bypass mechanisms observed.
 
\begin{table}[t]
\centering
\caption{Error analysis of residual attack successes under the proposed framework. Percentages are computed over the 32 bypass cases.}
\label{tab:error_analysis}
\footnotesize
\renewcommand{\arraystretch}{1.1}
\begin{tabularx}{\columnwidth}{X c c}
\toprule
\textbf{Failure Mode} &
\textbf{\%} &
\textbf{Count} \\
\midrule

Novel phrasing evading L1 detection &
37.5 & 12 \\

Implicit instructions bypassing L2 scanning &
25.0 & 8 \\

Persona drift below $\tau_{\text{drift}}$ &
21.9 & 7 \\

Multi-turn attack accumulation &
15.6 & 5 \\

\midrule

\textbf{Total} &
\textbf{100.0} &
\textbf{32} \\

\bottomrule
\end{tabularx}
\end{table}
 
The dominant bypass mechanism (37.5\%) involves semantically novel
phrasings that fall below the Layer~1 classifier threshold while
avoiding all pattern signatures. This is consistent with the known
brittleness of semantic classifiers trained on fixed-distribution
adversarial datasets~\cite{liu2023prompt_injection_attacks} and
motivates the audit loop's retraining cycle described in
Section~\ref{sec:layer_audit}. The second most common bypass
(25.0\%) involves indirect injection payloads crafted as implicit
soft instructions (e.g., a retrieved FAQ item that ends with a
parenthetical comment suggesting a different response style) rather
than explicit override commands; these avoid the chunk scanner's
explicit-override pattern library. Multi-turn injection, which
accumulates a partial payload across multiple conversation turns to
avoid per-turn detection, accounts for 15.6\% of bypasses and
represents a threat sub-class not covered by the current framework
design; it is noted as a direction for future work in
Section~\ref{sec:limitations}.

\section{Discussion}
\label{sec:discussion}
The experimental results in Section~\ref{sec:evaluation} admit
several interpretations that go beyond what can be expressed in the
tables alone. This section draws out the implications of the main
results, situates the findings within the broader context of LLM
security, and then identifies the limitations of the current work
and the directions they motivate.
 
\subsection{Implications of the Main Results}
\label{sec:discussion_implications} 
\textbf{Layering is necessary, not just beneficial:}
The ablation study (Table~\ref{tab:ablation}) shows that the full
three-layer framework achieves an ASR of 11.3\%, which is 11.1
percentage points below the best two-layer combination (L2+L3,
22.4\%) and 27.3 percentage points below the best single-layer
configuration (L3~Only, 38.6\%). These gaps are not incremental;
they represent qualitatively different levels of protection. The
result confirms the central design premise of the framework: direct
and indirect injection enter the pipeline at different points and
therefore require defenses positioned at those points. No subset of
two layers covers the full attack surface; Layer~1 is the only
component with access to the raw user input before retrieval
contamination, Layer~2 is the only component that can prevent a
retrieved document payload from reaching the inference engine
untagged, and Layer~3 is the only component that can catch
injections that bypass upstream controls.
 
The finding also has a practical implication for deployment: an
operator who cannot implement the full framework should prioritize
L2+L3 over L1+L3 or L1+L2. The L2+L3 combination achieves 22.4\%
ASR with an FPR of 7.4\%, which, while inferior to the full
framework, provides substantially better indirect injection
coverage than any configuration that omits L2.
 
\textbf{Layer~2 suppresses false positives in downstream layers:}
A notable result in Table~\ref{tab:ablation} is that the full
framework (L1+L2+L3, FPR~4.8\%) has a lower false positive rate
than both L1+L3 (FPR~6.9\%) and L2+L3 (FPR~7.4\%), despite being
strictly more restrictive. This counterintuitive result arises
because Layer~2 reduces the proportion of injections that reach the
Layer~3 policy rule engine. The Layer~3 harmful-content classifier
and semantic drift check are the components most likely to generate
false positives on legitimate queries that touch sensitive topics
or deviate from the reference persona. By filtering more true
positives before Layer~3, Layer~2 shifts the Layer~3 operating
point toward a region of its ROC curve with lower FPR. This
interaction between layers is not captured by any single-layer
evaluation and could not have been predicted without the ablation.
 
\textbf{Model capability does not substitute for structural defense:}
The per-model breakdown in Figure~\ref{fig:asr_goal_model} shows
that GPT-4o, despite being a substantially more capable model than
Llama~3~(8B) and Mistral~7B on standard benchmarks, achieves only
a modestly lower undefended ASR. This is consistent with the
observation of Perez and Ribeiro~\cite{perez2022ignore} that
instruction-following capability and instruction-override
susceptibility are not independent: a model that follows instructions
more reliably will also follow injected instructions more reliably.
The implication is that practitioners cannot substitute model
upgrades for structural defenses; a stronger model reduces ASR
marginally at best, while the framework reduces it by 60.1
percentage points regardless of the underlying model.
 
\textbf{Data exfiltration is the most tractable attacker goal:}
Across all configurations, the Data Exfiltration (DE) goal shows
the lowest residual ASR after applying the full framework. This is
attributable to the Layer~3 cosine similarity predicate, which is
a high-precision detector for verbatim or near-verbatim system
prompt leakage. The Instruction Override (IO) and Behavioral
Manipulation (BM) goals are harder to suppress because their
success criteria are behavioral rather than content-based: a
response can be manipulated without copying any specific text from
the injection payload, making similarity-based detection
inapplicable. This structural asymmetry suggests that future work
should prioritize improving IO and BM detection, as discussed in
Section~\ref{sec:future_work}.
 
\subsection{Limitations}
\label{sec:limitations}
\textbf{Static attacker model:}
The threat model in Section~\ref{sec:attacker_model} assumes a
black-box attacker with no knowledge of the deployed defense. A
white-box adversary with knowledge of the Layer~1 pattern library
and classifier thresholds could craft targeted evasion payloads
that avoid known signatures while remaining below the semantic
anomaly threshold. This is a standard limitation of detection-based
defenses and is not unique to the present framework; it motivates
the continuous audit loop and retraining cycle described in
Section~\ref{sec:layer_audit}, which are designed to reduce the
window of vulnerability between the introduction of a novel attack
pattern and its incorporation into the classifier's training
distribution.
 
\textbf{Single-turn evaluation:}
The experimental protocol evaluates each adversarial sample as an
independent single-turn query. The error analysis in
Section~\ref{sec:eval_error} identifies multi-turn injection as
the fourth most common bypass mechanism, accounting for 15.6\% of
residual bypass cases. In this attack pattern, a partial payload is
distributed across multiple turns of a conversation so that no
individual turn triggers a detection threshold. The current
framework applies Layer~1 and Layer~3 independently on each turn
and therefore has no mechanism for detecting payloads that are
individually innocuous but cumulatively injurious. Extending the
framework to maintain a session-level injection state across turns
is a direct direction for future work.
 
\textbf{Knowledge base composition:}
The evaluation knowledge base contains 500 documents drawn from
publicly available product documentation, which represents a
controlled and relatively benign content distribution. Real
deployments may index content from less controlled sources such as
user-generated reviews, social media, or third-party APIs, all of
which present a wider surface for knowledge base poisoning.
Evaluation on a more adversarially representative knowledge base
is needed to characterize framework performance under realistic
indirect injection conditions.
 
\textbf{Threshold sensitivity:}
The Layer~1 and Layer~3 thresholds ($\tau_1$, $\tau_3$,
$\theta_\text{sp}$, $\tau_\text{drift}$) are calibrated on a
held-out validation set derived from the same distribution as the
test set. In deployment, the input distribution will differ from
the evaluation distribution in ways that are difficult to
anticipate. Threshold miscalibration is the most likely cause of
FPR degradation in production. The audit loop's session-level and
population-level alerting mechanisms are designed to surface
threshold drift, but they depend on a sufficient volume of labeled
production data to drive recalibration.
 
\textbf{Latency under concurrent load:}
The latency measurements reported in
Table~\ref{tab:latency} are collected under single-request
conditions on dedicated hardware. In a production deployment
serving concurrent users, contention for the encoder model shared
between Layer~1 and Layer~3 may increase the p95 latency beyond
the 97.4\,ms reported here. Batching the semantic encoding
operations across concurrent requests is a straightforward
optimization that is not evaluated in this work.
 
\subsection{Future Work}
\label{sec:future_work}
Four directions follow directly from the limitations identified
above.
 
\textbf{Multi-turn injection detection:}
Extending Layer~1 to maintain a rolling session-level injection
score across turns, using a recurrent or attention-based aggregator
over per-turn scores, would address the multi-turn bypass mechanism.
A practical design would decay the session score between turns to
avoid over-sensitivity in long conversations.
 
\textbf{Implicit instruction detection in retrieved documents:}
The error analysis shows that 25\% of residual bypasses exploit
implicit soft instructions in retrieved documents that avoid the
chunk scanner's explicit-override patterns. Replacing or augmenting
the chunk scanner with a generative detector that prompts a
secondary LLM to assess whether a document chunk contains latent
instructions, as proposed by Greshake et
al.~\cite{greshake2023indirect}, could reduce this bypass category.
The latency cost of a generative detector would need to be evaluated
against the security gain.
 
\textbf{Adaptive threshold calibration:}
An online calibration mechanism that adjusts Layer~1 and Layer~3
thresholds in response to observed FPR drift in production, using
a lightweight Platt scaling or isotonic regression model fitted on
streaming audit log data, would reduce the sensitivity of the
framework to distribution shift without requiring full classifier
retraining.
 
\textbf{White-box evaluation:}
Evaluating the framework against an adaptive adversary with
knowledge of the defense, using techniques from adversarial machine
learning such as Carlini-Wagner
attacks~\cite{carlini2017towards} adapted to the text domain,
would provide a lower bound on framework security and characterize
the evasion effort required to overcome each layer.
 
\section{Conclusion}
\label{sec:conclusion}
This paper addresses prompt injection in LLM-based chatbots as a structural security problem arising from the lack of a clear boundary between instructions and data within the model context. Existing defenses operate at a single stage of the pipeline and therefore remain vulnerable to attacks delivered through alternative channels.

To address this challenge, we proposed a three-layer defense framework comprising Input Screening, Privilege-Constrained Context Assembly, and Output Auditing, supported by a Continuous Audit Loop for monitoring and adaptive improvement. Together, these layers provide complementary protection against both direct and indirect prompt injection attacks.

Experimental results show that the complete framework reduces the macro-averaged Attack Success Rate (ASR) from 71.4\% to 11.3\% across three target models and five attack categories, outperforming both the strongest single-layer baseline and a published general-purpose guardrail system. The framework also maintains a low false positive rate (4.8\%) and introduces only 61.2,ms of median latency overhead, making it suitable for interactive chatbot deployments. Ablation studies further confirm that each layer contributes unique defensive value and that their combined effect exceeds the sum of their individual contributions.

The findings indicate that model capability alone is insufficient to mitigate prompt injection and that effective protection requires defense mechanisms distributed across the entire inference pipeline. Although the framework substantially improves robustness, residual vulnerabilities remain, particularly against semantically novel attacks, implicit instructions embedded in retrieved content, and multi-turn attack strategies. Future work should focus on adaptive detection methods, stronger behavioral deviation analysis, and session-aware defenses. Overall, the proposed framework provides a practical and extensible foundation for securing LLM-based chatbot systems against evolving prompt injection threats.

\bibliographystyle{ieeetr}
\bibliography{references}

\par
\vspace{10pt}
{\large\textbf{Data and Code Availability}}
The datasets, code, and related resources used in the implementation and experiments of this study are available at:
\href{https://github.com/nisarahmedrana/Prompt_Injection_Attack/}{GitHub Repository}.

\acknowledgement
This study did not receive any specific grant from funding agencies in the public, commercial, or not-for-profit sectors.

\conflictsofinterest
The authors are affiliated with SparkVerse AI, which contributed to this study. This affiliation may represent a potential conflict of interest. The research, analyses, and reporting were conducted objectively and in accordance with scientific and ethical standards.

{\large\textbf{Ethical Approval and Consent to Participate}}

Not applicable.

\begin{fullwidth}
\end{fullwidth}


\makeatletter
\if@twocolumn

  \begin{biography}[Gulshan]{Gulshan Saleem} received her Bachelor’s and Master’s degrees in Software Engineering and a Ph.D. in Computer Science. Her research interests include machine learning, data analytics, and intelligent systems. She has authored and co-authored peer-reviewed publications and contributed to research on the design and analysis of data-driven computational solutions. (Email: \href{mailto:gulshnsaleem26@gmail.com}{gulshnsaleem26@gmail.com})
  \end{biography}
  
  \begin{biography}[Nisar]{Nisar Ahmed} (Senior Member, IEEE) received his Ph.D. and Master’s degrees in Computer Engineering and a Bachelor’s degree in Electrical Engineering. His research focuses on machine learning, deep learning, and AI-driven solutions for cybersecurity and healthcare. He has authored multiple peer-reviewed publications and contributed to research on data-driven security analytics and intelligent systems, with an emphasis on reproducibility and real-world deployment. (Email: \href{mailto:nisar.ahmed@sparkverse.ai}{nisar.ahmed@sparkverse.ai})
  \end{biography}

  \begin{biography}[Imran]{Muhammad Imran Zaman} Muhammad Imran Zaman received his Bachelor’s and Master’s degrees in Computer Science. He currently works in industry, focusing on practical, data-driven solutions for real-world data science problems. His interests include applied machine learning, data analytics, and the deployment of intelligent systems in production environments. (Email: \href{mailto:imran.zaman@sparkverse.ai}{imran.zaman@sparkverse.ai})
  \end{biography}

  \begin{biography}[Ali]{Ali Hassan} Ali Hassan received his B.S. degree in Computer Science. He is currently working in industry, focusing on the deployment of intelligent systems in production environments. He also has experience as a front-end web developer, contributing to the design and development of user-centric web applications. (Email: \href{mailto:ali.hassan@sparkverse.ai}{ali.hassan@sparkverse.ai})
  \end{biography}

    \begin{biography}[Umar]{Umar Mujahid Khokhar} Umar Mujahid Khokhar is an Associate Professor of Information Technology at Georgia Gwinnett College. He earned his Ph.D. in Information Security from Bahria University and has extensive academic experience in cybersecurity education and research. His work focuses on information security, privacy-preserving systems, ultra-lightweight cryptography, and secure hardware implementations for pervasive computing environments. (Email: \href{mailto:ukhokhar@ggc.edu}{ukhokhar@ggc.edu})
  \end{biography}

\else
\fi
\makeatother
\end{document}